\documentstyle[preprint,aps,psfig]{revtex}

\def\lsim{\raise0.3ex\hbox{$\;<$\kern-0.75em\raise-1.1ex
\hbox{$\sim\;$}}}
\def\gsim{\raise0.3ex\hbox{$\;>$\kern-0.75em\raise-1.1ex
\hbox{$\sim\;$}}}
\begin{document}

\draft

\preprint{}

\title{CP Violation vs. Matter Effect in Long-Baseline Neutrino 
Oscillation Experiments} 

\author{Hisakazu Minakata$^1$ and Hiroshi Nunokawa$^2$}

\address{\qquad \\ {$^1$}Department of Physics,
Tokyo Metropolitan University \\
Minami-Osawa, Hachioji, Tokyo 192-03, Japan\\
\qquad\\{$^2$}Instituto de F\'isica Corpuscular - C.S.I.C. \\
Department of F\'isica Te\`orica, Universitat de Val\'encia, \\
46100 Burjassot, Valencia, Spain\\}

\date{April, 1997}
\preprint{
\parbox{5cm}{
TMUP-HEL-9704\\
FTUV/97-21\\
IFIC/97-21\\
hep-ph/9705208\\
Revised December 1997\\
}}
\maketitle

\begin{abstract}
We investigate, within the framework of three generations of neutrinos, 
the effects of $CP$ violation in long-baseline neutrino oscillation 
experiments. We aim at illuminating the global feature of the interplay 
between genuine effect due to the $CP$ violating phase and a fake one 
due to the earth matter effect. To this goal, we develop a formalism 
based on the adiabatic approximation and perturbative treatment of the 
matter effect which allows us to obtain approximate analytic expressions 
of the oscillation probabilities. 
We present an order-of-magnitude estimation and a detailed numerical 
computation of the absolute and the relative magnitudes of the $CP$ 
violations under the mass hierarchy suggested by the atmospheric 
neutrino anomaly and the cosmological dark matter. We find that the 
genuine $CP$ violating effect is at most $\sim $ 1 \%, and the matter 
effect dominates over the intrinsic $CP$ violation only in a region 
of parameters where the oscillation probability of 
$\nu_\mu\rightarrow\nu_e$ is large. 

\end{abstract}
\vskip 0.5cm
\pacs{14.60.Pq, 25.30.Pt, 95.35.+d}

\section{Introduction}
The origin of $CP$ violation is still a mystery in particle physics. 
Unlike in the quark sector even the very existence of $CP$ violation 
is not known in the lepton sector. Recent advances in neutrino 
observations mainly of astrophysical origins strongly suggest the 
existence of tiny neutrino masses \cite {solar,atmospheric}. If this 
is the case nature would admit the Cabibbo-Kobayashi-Maskawa (CKM) 
\cite {CKM} type flavor mixing also in the lepton sector which allows 
us to have $CP$ violating phases with three (or more) generation of 
leptons. If revealed, it should give us important insight into our 
understanding of fundamental structure of matter. Moreover, there is 
an intriguing suggestion \cite {FY} that $CP$ violation in the lepton 
sector might be an indispensable ingredient in producing the baryon 
asymmetry in the universe. 

It has been known since long time ago \cite {Cabibbo} that the 
effect of $CP$ violating phase can in principle be observable in 
neutrino oscillation experiments. The other earlier references 
include \cite {Barger,Pakvasa,Bilenky}. As we will 
recapitulate below the particle-antiparticle difference between 
the oscillation probabilities, 
$\Delta P(\nu_{\beta}\rightarrow\nu_{\alpha}) 
\equiv P(\nu_{\beta}\rightarrow\nu_{\alpha}) 
- P(\bar{\nu_{\beta}}\rightarrow\bar{\nu_{\alpha}})$, in vacuum 
is characterized \cite {Barger,Pakvasa} by the leptonic analogue 
of the Jarlskog factor, the unique and phase-convention 
independent measure for $CP$ violation \cite {Jarlskog}. 

Recently, measuring leptonic $CP$ violation in the long-baseline 
neutrino oscillation experiments  \cite {KEKSK,MINOS,ICARUS} 
received considerable amount of attention in the literature 
\cite {Tanimoto,AS}. The potential obstacle in measuring $CP$ 
violation in the long-baseline experiment is the matter effect in 
the earth. It is well known that the matter effect acts differently 
in propagation of $\nu$ and $\bar\nu$ in matter; it gives rise to 
the index of refraction which differs in sign between $\nu$ and 
$\bar\nu$ \cite {Wolfenstein}.  Then, the $\nu-\bar\nu$ difference 
$\Delta P$ of the oscillation probabilities is inevitably contaminated 
by the matter effect \cite {matter}. 
In fact, it is known by numerical computation 
that the matter effect is overwhelming over the genuine $CP$ violating 
effect at certain values of the mixing parameters in the 
$\nu_\mu\rightarrow\nu_e$ channel \cite {Tanimoto,AS}. 
But it appears to the authors that we still lack understanding of 
the over-all features of the relationship between the $CP$ violations 
due to the matter and to the $CP$ violating phases. 

It is the purpose of the present paper to illuminate global structure 
of the interplay between the matter and the genuine $CP$ violating 
effects in long-baseline experiments. To this goal we develop a 
formalism by which we can derive approximate analytic expressions of 
oscillation probabilities. These analytic formulas will allow us to have 
global view of the features of $CP$ violation in neutrino oscillation 
experiments. Our formalism is based on the adiabatic approximation and 
takes into account the matter effect in a perturbative way. It also 
enjoys further simplification due to the presumed hierarchy of neutrino 
masses which will be explained below.   

This is the first in a series of papers to investigate the same problem 
in various neutrino mass hierarchies. 
In this paper, we assume that the atmospheric neutrino anomaly, which 
is observed by the Kamiokande, IMB, and the Soudan 2 experiments
\cite {atmospheric}, can be interpreted as the evidence for neutrino 
oscillations whose relevant mass scale is 
$\Delta m^2 \sim 10^{-2}$ eV$^2$ or larger. 
(We note, however, that the two experiments do not
observe the atmospheric neutrino anomaly \cite {nusex}.) 
It is a natural assumption because 
it is the very motivation for planning the long-baseline experiments.
The restriction leads to the hierarchy between the matter potential
and the mass differences, which allows us to treat the matter effect
perturbatively. Namely, we derive closed-form analytic expressions of 
the neutrino-antineutrino difference between oscillation probabilities 
which is generally valid under the adiabatic approximation and a 
first-order perturbative treatment of the matter effect.

We restrict ourselves into the case of neutrino mass hierarchy motivated 
by the cosmological hot dark matter \cite{hcdm} in our analysis of the 
features of $CP$ violation in this paper. 
The assumption of the dark matter scale mass difference allows us to 
utilize the strong constraints on the mixing parameters deduced from 
the terrestrial experiments \cite{raconstr}. 
It is the reason why we decided to investigate this case first; Because 
of the constraints we are able to estimate the magnitude of CP violation 
in a less ambiguous way.

We give a few ward on the problem if the mass hierarchy we examine in 
this paper can account for the other known hints for the neutrino 
masses. They include the solar neutrino deficit \cite {solar} and the 
LSND experiments \cite {LSND}. Let us first discuss the solar neutrinos. 
It is well known that for this scheme the capability of explaining the 
gross deficit of the solar neutrino flux depends upon the parameter 
regions. In the region (A) 
(see Eq. (\ref{eqn:region}) for the definitions of these regions) 
the global deficit of the flux can be accommodated in a manner of 
Acker and Pakvasa \cite {AP}. In the region (B) there is no way of 
having a gross deficit of the solar neutrino flux within the framework 
of three-flavor mixing. Therefore, we must introduce some other 
ingredients such as sterile neutrinos to achieve the deficit in the 
region (B). 

The results of the LSND experiments can be easily accommodated in the 
present framework. The LSND data allow the mass scales which is 
consistent with hot dark matter. If smaller values of $\Delta m^2$ is 
preferred for consistency with other experiments it may be necessary 
to relax the constraints on the mass scale required by the mixed dark 
matter cosmology \cite {hcdm}. If the simultaneous consistency with 
the solar neutrino and the LSND data is demanded there is no solution 
within our present framework. 

Within the restrictions to the mass hierarchies hinted by the dark 
matter and the atmospheric neutrinos, 
we will try to answer the following questions: \\
(1) Are there any channels which are much less contaminated by the 
matter effect? \\
(2) In which parameter regions do we expect to have maximal $CP$ 
violation and how large is its size? 

It is known that the matter effect in the oscillation probability 
in long-baseline experiments is not very large, at most a few to 
several $\%$. One might then feel strange that the matter effect 
is dominant in certain channel. The point is that what we are
dealing with is not the oscillation probability itself but the
difference between the neutrino and the antineutrino oscillation
probabilities. The matter effect can give rise to a dominant effect
in such $\nu-\bar\nu$ difference.

This investigation was motivated by a question asked by a 
long-baseline neutrino experimentalist \cite {Ayres}; 
``Is the matter effect contamination small in 
$\nu_\mu\rightarrow\nu_\tau$ channel?''
This is the interesting question for two reasons, one theoretical 
and one experimental. Experimentally it is a very relevant question 
because they are planning to do long-baseline neutrino experiments 
in the appearance channel, $\nu_\mu\rightarrow\nu_\tau$. 
It is also of interest from the theoretical point of view; 
In the conventional treatment 
of ``optional'' two-flavor mixing favored by experimentalists the 
$\nu_\mu\rightarrow\nu_\tau$ channel might be free from the matter 
effect because there is no $\mu$'s and $\tau$'s in the earth. 
On the other hand, $CP$ violation is the genuine three-flavor mixing 
effect which cannot occur in two-flavor mixing framework. 
Therefore, the question still remains whether the matter effect is 
small in the $\nu_\mu\rightarrow\nu_\tau$ channel, as correctly 
raised by the experimentalist. 

In Sec. II, we review some basic facts on $CP$ violation in vacuum 
in the context of neutrino oscillation experiments. We also summarize 
the mass patterns of neutrinos which we use in our analysis in this 
paper.  In Sec. III, we set up our formalism based on the adiabatic 
approximation. In Sec. IV, we develop a framework for perturbative 
treatment of the matter effect which is applicable to the long-baseline 
neutrino experiments. In Sec. V, we derive the approximate analytic 
formulas for the neutrino-antineutrino difference in oscillation 
probabilities by taking account of the matter effect to first-order 
in perturbation theory. 
In Sec. VI, we use our analytic formula to illuminate global features 
of the competing two effects producing $CP$ violation, the genuine 
effect due to $CP$ violating phase and a fake one due to the matter 
effect. 
In Sec. VII, we give the results of our detailed numerical computation 
of the $CP$ violation to confirm the qualitative understanding of the 
structure of the coexisting two effects obtained in Sec. VI. 
The last section VIII is devoted to the conclusion. 
In Appendix we confirm that the adiabatic approximation is in fact 
a very good approximation for matter density profiles relevant for 
the long-baseline experiments. 

\section{CP Violation in Vacuum and Neutrino Mass Spectrum Motivated
by Dark Matter and Atmospheric Neutrino Anomaly}

In this section we first review briefly the $CP$ violation in neutrino
oscillations in vacuum. We then describe the neutrino mass spectrum 
which we consider in this paper. It is intended for convenience for 
the readers who are not familiar to the subject and precedes the 
following three sections in which we develop our analytic framework 
usable for analyses with more general neutrino mass patterns. 

We work with the three-flavor mixing scheme of neutrinos and 
introduce the flavor mixing matrix $U$ as
\begin{equation}
\nu_{\alpha}=U_{\alpha i}\nu_i,
\end{equation}
where $\nu_{\alpha}(\alpha=e,\mu, \tau)$ and $\nu_i(i=1,2,3)$
stand for the gauge and the mass eigenstates, respectively.
Then in vacuum the direct measure of $CP$ violation can be written
\cite{Barger,Pakvasa} as 
\begin{equation}
\Delta P(\nu_{\beta}\rightarrow\nu_{\alpha}) \equiv
P(\nu_{\beta}\rightarrow\nu_{\alpha}) -
P(\bar{\nu_{\beta}}\rightarrow\bar{\nu_{\alpha}})
= -4J_{\beta\alpha}(\sin\Delta_{12}+\sin\Delta_{23}+\sin\Delta_{31}),
\label{eqn:deltap}
\end{equation}
where 
\begin{equation}
\Delta_{ij}\equiv \frac{(m_j^2-m_i^2)L}{2E}, 
\end{equation}
with $m_i (i=1,2,3)$ being the mass of $i$-th neutrino, $E$ the neutrino 
energy and $L$ the distance from the neutrino source.
We have used in (\ref {eqn:deltap}) the Jarlskog parameter
\begin{equation}
J_{\beta\alpha}\equiv \mbox{Im}
[U_{\alpha 1}U_{\alpha 2}^*U_{\beta 1}^*U_{\beta 2}]
\end{equation}
which is unique, up to the sign, in three-flavor neutrinos. 

We will use the following form of the neutrino mixing matrix
\begin{equation}
U=\left[
\begin{array}{ccc}
c_{12}c_{13} & s_{12}c_{13} &   s_{13}e^{-i\delta}\nonumber\\
-s_{12}c_{23}-c_{12}s_{23}s_{13}e^{i\delta} &
c_{12}c_{23}-s_{12}s_{23}s_{13}e^{i\delta} & s_{23}c_{13}\nonumber\\
s_{12}s_{23}-c_{12}c_{23}s_{13}e^{i\delta} &
-c_{12}s_{23}-s_{12}c_{23}s_{13}e^{i\delta} & c_{23}c_{13}\nonumber\\
\end{array}
\right],
\label{eqn:ckmmatrix}
\end{equation}
which
is identical with the standard CKM matrix for quarks. Here
$c_{ij}\equiv\cos\theta_{ij}$ and $s_{ij}\equiv\sin\theta_{ij}$
where $(i,j)$ = (1,2), (2,3) and (1,3).
With this parametrization $J_{\beta\alpha}$ can be expressed as
\begin{equation}
J_{\beta\alpha} = \pm J, \ \ \
J \equiv c_{12}s_{12}c_{23}s_{23}c_{13}^2s_{13}\sin\delta,
\end{equation}
where $+$ sign is for cyclic permutations, i.e.,
$(\alpha,\beta)=(e,\mu),(\mu,\tau),(\tau,e)$ and $-$
is for anti-cyclic ones.

Let us come to the mass hierarchy which will be used in our analysis 
presented in Secs. VI and VII. 
We work with the mass hierarchy 
\begin{eqnarray}
|\Delta m^2| \sim 10^{-2} \mbox{eV}^2, \ \
\Delta m^2  &\equiv & m_2^2-m_1^2,
\label{eqn:assatm}\\
|\Delta M^2| \gsim \mbox{a few\ eV}^2, \ \
\Delta M^2  &\equiv & m_3^2-m_1^2 \simeq  m_3^2-m_2^2,
\label{eqn:assdark}
\end{eqnarray}
where the first is suggested by the atmospheric neutrino anomaly 
\cite {atmospheric}, and the second is motivated by the hot and 
the cold dark matter cosmology \cite{hcdm}.

The Kamiokande, IMB, and the Soudan 2 experiments observed 
a 30-40 \% deficit in the double ratio
$(\frac {\nu_{\mu}}{\nu_{e}})_{observed}
/(\frac {\nu_{\mu}}{\nu_{e}})_{expected}$ 
\cite{atmospheric}. 
A natural interpretation of the anomaly is due to the neutrino 
oscillation. In particular, the multi-GeV data from the Kamiokande 
experiments indicates that the double ratio has a zenith-angle 
dependence which is quite consistent with the interpretation of 
the anomaly as the evidence for neutrino oscillations with the 
mass difference $\Delta m^2 \sim 10^{-2}$ eV$^2$. 

The hot and the cold dark matter cosmology is one of the viable models 
of the structure formation in the universe \cite{hcdm}. The hot dark 
matter is the indispensable ingredient in the scenario by which the 
magnitude of 
density fluctuation normalized by COBE data can be made consistent 
with that of small scales determined by correlations between galaxies 
and clusters \cite {WS}. Neutrinos of masses 2-20 eV are the natural 
candidate for the hot dark matter. In fact, it is the only known 
particles among the candidates for particle dark matter.

The advantage of the dark matter motivated mass hierarchy
(\ref{eqn:assdark}) is that the mixing parameters are subject to
the powerful constraints that comes from the reactor and
the accelerator experiments \cite{raconstr} and we can draw 
a clear answer to the question on how large is the magnitude of 
$CP$ violation.

Under the assumption (\ref{eqn:assdark}) the oscillations due to 
the larger mass squared difference $\Delta M^2 $ are averaged out 
in long-baseline experiments and we have,
\begin{equation}
\Delta P(\nu_\beta \rightarrow \nu_\alpha) \sim
- 4J_{\alpha\beta}\sin\left(\frac{\Delta m^2 L}{2E}\right)
\end{equation}
Since 
$\left(\frac{\Delta m^2 L}{2E}\right) = 2.5  
\left(\frac{\Delta m^2}{10^{-2}\mbox{eV}^2}\right)
\left(\frac{L}{100 \mbox{km}}\right)
\left(\frac{E}{1\mbox{GeV}}\right)^{-1}$
can be of order unity in long-baseline experiments, 
the $CP$ violation $\Delta P_{\beta\alpha} $ 
can be approximated as $\sim -4J_{\alpha\beta}$.

Let us estimate how large $CP$ violation can be for our choice
of mass hierarchy (\ref{eqn:assatm}) and (\ref{eqn:assdark}).
In Fig. 1 we present equal-$J$ contours on 
$\tan^2\theta_{13}-\tan^2\theta_{23}$ plane with use of the 
parameters $\theta_{12}=\pi/4$ and $\delta= \pi/2$ which 
maximize $J$. In the same figure we also plot the regions of 
parameters excluded by the reactor and accelerator experiments 
for the case $|\Delta M^2| = 5$ eV$^2$ obtained in Ref. \cite{Fogli}.
We notice from Fig. 1 that allowed parameters are restricted
into three separate regions.
\begin{eqnarray}
&(\mbox{A})& \  \  \mbox{small}\  -s_{13}\ \
\mbox{and\ \ small}\  \  -s_{23}\nonumber   \\ 
\label{eqn:region}
&(\mbox{B})& \  \  \mbox{large}\  -s_{13}\ \  
\mbox{and\ \ arbitrary}\  \  -s_{23} \\
&(\mbox{C})& \  \  \mbox{small}\  -s_{13}\ \  
\mbox{and\ \ large}\  \  -s_{23} \nonumber
\end{eqnarray}
The constraints from the reactor and accelerator experiments are
so strong that one neutrino flavor almost decouples with the remaining 
two. Namely, the $\nu_\tau$, $\nu_e$ and $\nu_\mu$ almost decouple
with the other two flavors in the regions (A), (B) and (C),
respectively. The remaining two neutrinos can be strongly mixed 
with each other in each allowed region. We observe that the value 
of $4J$ is at most $\simeq 0.01$ in the regions (A) and (B) and 
$\simeq 0.001$ in the region (C).

In Fig. 2 we show the neutrino mass spectrum which is realized in 
the region (A) and (B). We do not consider in this paper the region 
(C) since the atmospheric neutrino anomaly cannot be accounted for 
and moreover $CP$ violation is very small in this region.
There exist two different mass patterns which can be realized 
in each parameter regions. Depending upon the sign of $\Delta M^2$ the 
decoupled state $\nu_3$ can be the heaviest, (A-1) or (B-1), or 
the lightest, (A-2) or (B-2). 
The case (A-1) is theoretically appealing because of the seesaw 
mechanism \cite{seesaw} and from the observed mass hierarchy among 
charged leptons.

\section{Adiabatic Neutrino Evolution in Matter}

We discuss the neutrino propagation in the earth matter within 
the framework of three-flavor mixing of neutrinos. It is worth to 
note that when we discuss the long-baseline experiments whose 
baseline distances less than 1000 km the adiabatic approximation 
is a very good approximation, as we demonstrate in Appendix.  
It is because the neutrino beam only passes through thin layer 
of the continental structure closest to the earth surface in which 
the matter density is approximately constant, $\simeq$ 2.72 g/cm$^3$ 
\cite {Stacey}. 
By virtue of this fact we can derive a general closed form 
expression of $\Delta P$, the $\nu - \bar\nu$ difference in 
oscillation probabilities. 

The evolution equation of neutrinos (and antineutrinos) can be written 
in terms of the gauge eigenstate as 
\begin{equation}
i\frac{d}{dx} 
\left[
\begin{array}{c}
\nu_e \\ \nu_\mu \\ \nu_\tau
\end{array}
\right] 
=
H(x)
\left[
\begin{array}{c}
\nu_e \\ \nu_\mu \\ \nu_\tau
\end{array}
\right],
\end{equation}
where the Hamiltonian $H(x)$ is given by 
\begin{equation}
H=
U \left[
\begin{array}{ccc}
m_1^2/2E & 0 & 0 \\
0 & m_2^2 /2E & 0 \\
0 & 0 & m_3^2/2E 
\end{array}
\right] U^{+}
+
\left[
\begin{array}{ccc}
a(x) & 0 & 0 \\
0 & 0 & 0 \\
0 & 0 & 0
\end{array}
\right],
\label{eqn:hamiltonian}
\end{equation}
where $x$ is the position along the neutrino trajectory.
Here $a(x)$ represents the matter effect and has the form 
\begin{equation}
a(x)=\pm \sqrt 2 G_F N_e(x)
\end{equation}
where $G_F$ is the Fermi constant, $N_e(x)$ is the electron number 
density at $x$ in the earth, and the + and $-$ signs are for neutrinos 
and antineutrinos, respectively. 

We introduce the unitary transformation $V(x)$ which diagonalize 
$H(x)$ locally, i.e., at each point $x$ of neutrino trajectory; 
\begin{equation}
V^{+}(x)H(x)V(x) = H_d(x)
\end{equation}
and parametrize the diagonalized Hamiltonian as 
$H_d(x)= \mbox{diag.} [h_1(x),h_2(x),h_3(x)]$. 
We then define the matter-mass eigenstate, mass eigenstate basis 
in matter, as
\begin{equation}
\left[
\begin{array}{c}
\nu_{m_1} \\ \nu_{m_2} \\ \nu_{m_3}
\end{array}
\right] 
=
V^{+}(x)
\left[
\begin{array}{c}
\nu_e \\ \nu_\mu \\ \nu_\tau
\end{array}
\right],
\end{equation}

The neutrino evolution equation in terms of the matter-mass eigenstate 
takes the form 
\begin{equation}
i\frac{d}{dx} \left[
\begin{array}{c}
\nu_{m_1} \\ \nu_{m_2} \\ \nu_{m_3}
\end{array}
\right] =
H_d(x)
\left[
\begin{array}{c}
\nu_{m_1} \\ \nu_{m_2} \\ \nu_{m_3}
\end{array}
\right] + 
i\frac{d}{dx}V^{+}(x)\cdot V(x)
\left[
\begin{array}{c}
\nu_{m_1} \\ \nu_{m_2} \\ \nu_{m_3}
\end{array}
\right].
\label{eqn:mattermass}
\end{equation}
The adiabatic approximation amounts to ignore the second term in 
the right-hand-side of eq. (\ref{eqn:mattermass}). We will return in 
Appendix to the question if it gives a really good approximation. 

Under the adiabatic approximation it is straightforward to solve 
the evolution equation as 
\begin{equation}
\nu_\alpha (x) = V_{\alpha i}(x)
\mbox{exp}\left\{-i \int_0^x dx' h_i(x')\right\}
V_{\beta i}^*(0) \nu_\beta (0).
\label{eqn:solution}
\end{equation}
We then obtain the probability of the neutrino oscillation 
$\nu_{\beta}\rightarrow\nu_{\alpha}$ where $\nu_{\beta}$ is 
created at $x=0$ and $\nu_{\alpha}$ is detected at $x=L$. 
It reads
\begin{eqnarray}
P(\nu_\beta \rightarrow \nu_\alpha) &=& 
\sum_{i,j}V_{\alpha i}(L)V_{\alpha j}^*(L)V_{\beta i}^*(0)V_{\beta j}(0)
\nonumber\\
&&\times \mbox{exp}\left\{-i \int_0^L dx [h_i(x)-h_j(x)]\right\}.
\end{eqnarray}

We assume in this paper an idealized situation where the matter 
densities at the production and the detection points are identical. 
This should give a good approximation to the real experimental 
situation because these points are either on or are very close to 
the earth surface. One way argue that one can take $V(0)=V(L)$ equal 
to the vacuum mixing matrix $U$ by saying that we design the experiment 
so that the production and the detection points of neutrinos on the 
earth surface, namely, in vacuum. We argue that even with such 
experimental condition it is the better approximation to take as 
$V(0)=V(L)$ the value of matter-mass mixing matrix with matter density 
$\sim$ 2.7 g/cm$^3$ at just below the earth surface. 
First of all, if we take the vacuum mixing matrix $U$ for $V(0)=V(L)$ 
we have to worry about the failure of the adiabaticity condition at 
the earth surface. The oscillation length of neutrinos is approximately 
given by 
\begin{equation}
L=2.5\left(\frac{E}{1\mbox{GeV}}\right)
\left(\frac{\Delta m^2}{1\mbox{eV}^2}\right)^{-1}\mbox{km}
\end{equation}
and is much longer than the decay tunnel or the detector hole for 
interesting regions of $\Delta m^2$. Therefore, the neutrinos do not know 
if they take off at the earth surface prior to the detection and would 
rather feel as if they remain in the earth matter at the point of 
detection. 

By separating the summation into $i=j$ and $i\neq j$ terms one can rewrite 
the expression of oscillation probability into the form 
\begin{eqnarray}
P(\nu_\beta \rightarrow \nu_\alpha \;; \delta, a) 
&=& 
-4\sum_{j>i} 
Re[V_{\alpha i}(a)V_{\alpha j}^*(a)V_{\beta i}^*(a)V_{\beta j}(a)]
\sin^2[\frac {1}{2}I_{ij}(a)]\nonumber\\
&&+2\sum_{j>i} 
\mbox{Im}[V_{\alpha i}(a)V_{\alpha j}^*(a)V_{\beta i}^*(a)V_{\beta j}(a)]
\sin^1[I_{ij}(a)],
\label{eqn:probability}
\end{eqnarray}
where 
\begin{equation}
I_{ij}(a)=\int_0^L dx [h_i(x)-h_j(x)].
\label{eqn:iij}
\end{equation}

\section{perturbative treatment of matter effect}

Now we intend to evaluate $\Delta P$ to first-order in matter 
perturbation theory. Let us first confirm that the perturbative 
treatment is reliable under the mass scale with which we are 
working. Since we are assuming the mass difference 
$\Delta m^2 \sim 10^{-2}$ eV$^2$ 
relevant for atmospheric neutrino anomaly we have, 
\begin{equation}
\frac{\Delta m^2}{E}=
10^{-11}\left(\frac{\Delta m^2}{10^{-2}\mbox{eV}^2}\right)
\left(\frac{E}{1\mbox{GeV}}\right)^{-1}
\mbox{eV}.
\end{equation}
On the other hand, we estimate the matter potential for the continental 
structure as, 
\begin{equation}
a =1.04 \times 10^{-13}
\left(\frac{\rho}{2.7 \mbox{g/cm}^3}\right)
\left(\frac{Y_e}{0.5}\right)
\mbox{eV},
\end{equation}
where $Y_e \equiv N_p/(N_p + N_n)$ is the electron fraction. 
Hence we see that the hierarchy between the energy scales,
$a\ll \frac{\Delta m^2}{E} \;\mbox{(atmospheric mass scale)}$,
holds. 

We use stationary state perturbation theory in accord with the slow 
variation of $a(x)$ on which the adiabatic approximation is based. 
It is convenient to take the vacuum mass eigenstate $\nu_i(i=1,2,3)$, 
which diagonalizes the unperturbed Hamiltonian, as the basis of matter 
perturbation theory. We denote the first and the second terms of the 
Hamiltonian (\ref{eqn:hamiltonian}) as $H_0$ and $H'$, respectively. 
We use the vacuum mass eigenstate $\nu_i$ defined by 
\begin{equation}
\nu_i(x) = e^{-ih_i^{(0)}x}\nu_i, 
\end{equation}
where $h_i^{(0)}=m_i^2/2E$, as the basis of matter perturbation theory. 

The Hamiltonians in this basis are 
$\tilde{H_0}=U^+H_0U$ and $\tilde{H'}=U^+H'U$, respectively. 
Since $(H')_{\alpha\beta} = a\delta_{\alpha e}\delta_{\beta e}$ 
we have
\begin{equation}
(\tilde{H'})_{ji}=aU_{ei}U^*_{ej}
\end{equation}
Then, the matter mass eigenstate can be expressed to first order in 
$a$ as
\begin{equation}
\nu_{mi} = \nu_i + \sum_{j\neq i} 
\frac{aU^*_{ei}U_{ej}}{h_i^{(0)}-h_j^{(0)}}
\nu_{j}
\end{equation}
It can be converted to the expression of vacuum mass eigenstate 
expressed by the matter mass eigenstate 
\begin{equation}
\nu_{i} = \nu_{mi} -  
\sum_{j\neq i}\frac{aU^*_{ei}U_{ej}}{h_i^{(0)}-h_j^{(0)}}
\nu_{mj}
\end{equation}
Since the flavor eigenstate has the simple relationship with the 
vacuum mass eigenstate as $\nu_{\alpha}=U_{\alpha i}\nu_i$ 
we obtain
\begin{equation}
\nu_{\alpha}=\sum_i \left[
U_{\alpha i}\delta_{ij}-\sum_{j \neq i} \frac{U_{\alpha i}U^*_{ei}U_{ej}}
{h_i^{(0)}-h_j^{(0)}} a\right]\nu_{mj}
\end{equation}
The matrix element of $V$ can be read off from this equation as
\begin{equation}
V_{\alpha i} = U_{\alpha i} - \sum_{j \neq i}
\frac{U_{\alpha j}U^*_{ej}U_{ei}}{h_j^{(0)}-h_i^{(0)}}a
\label{eqn:Vperturb}
\end{equation}
It is instructive to verify that $V_{\alpha i}$ in (\ref{eqn:Vperturb}) 
satisfies the unitarity to first order in $a$, as it should. 

We have obtained the matrix $V$ in the form $V=U+\delta V$, where 
$\delta V$ denotes the correction first order in $a$. Then we can write, 
schematically, $VVVV$ in (\ref{eqn:probability}) as 
\begin{equation}
VVVV=UUUU+UUU\delta V
\end{equation}
where the second term actually contains four pieces. 
It is important to know the symmetry property of these terms. 
It follows that 
\begin{equation}
\begin{array}{ccc}
\mbox{Re}(UUUU) & 
\raisebox{-1.8mm}{\overrightarrow{\delta \rightarrow -\delta}} & 
\mbox{Re}(UUU\delta V) \\
\mbox{Im}(UUUU) &
\raisebox{-1.8mm}{\overrightarrow{\delta \rightarrow -\delta}} & 
-\mbox{Im}(UUUU) \\
\mbox{Re}(UUU\delta V) &
\raisebox{-1.8mm}
{\overrightarrow{\delta \rightarrow -\delta, a \rightarrow -a}} &
-\mbox{Re}(UUU\delta V) \\ 
\mbox{Im}(UUU\delta V) &
\raisebox{-1.8mm}
{\overrightarrow{\delta \rightarrow -\delta, a \rightarrow -a}} &
\mbox{Im}(UUU\delta V). \\ 
\end{array}
\label{eqn:symmetry}
\end{equation}
They stem from the fact that $UUU\delta V$ is linear in $a$ and that 
the transformation $\delta \rightarrow -\delta$ is equivalent to 
$U \rightarrow U^*$. 

\section{CP Violation Effect in the Presence of Matter}

The neutrino-antineutrino difference can be written as
\begin{eqnarray}
\Delta P(\nu_{\beta}\rightarrow\nu_{\alpha})&\equiv&
P(\nu_{\beta}\rightarrow\nu_{\alpha})-P(\bar\nu_{\beta}-\bar\nu_{\alpha})
\nonumber\\
&=&P(\nu_{\beta}\rightarrow\nu_{\alpha}\;;\delta,a)
-P(\nu_{\beta}\rightarrow\nu_{\alpha}\;;-\delta,-a)
\label{eqn:deltaP0}
\end{eqnarray}
Thanks to the symmetry properties (\ref{eqn:symmetry}), we 
can express eq. (\ref{eqn:deltaP0}) as, 
\begin{eqnarray}
\Delta P(\nu_{\beta}\rightarrow\nu_{\alpha})=
 &-& 4 \sum_{j>i} \mbox{Re}(UUUU)_{\alpha\beta \: ;\:  ij}
\biggl[\sin^2\{\frac{1}{2}I_{ij}(a)\}-\sin^2\{\frac{1}{2}I_{ij}(-a)\}\biggr] 
\cr
&+&2 \sum_{j>i}\mbox{Im}(UUUU)_{\alpha\beta \: ;\: ij}
\biggl[\sin{I_{ij}(a)} + \sin{I_{ij}(-a)}\biggr] \cr
&-& 4 \sum_{j>i} \mbox{Re}(UUU \delta V)_{\alpha\beta \: ;\:  ij}
\biggl[\sin^2\{\frac{1}{2}I_{ij}(a)\}+\sin^2\{\frac{1}{2}I_{ij}(-a)\}\biggr] 
\cr
&+& 2 \sum_{j>i}\mbox{Im}(UUU\delta V)_{\alpha\beta \: ;\: ij}
\biggl[\sin{I_{ij}(a)} - \sin{I_{ij}(-a)}\biggr],
\label{eqn:deltaP1}
\end{eqnarray}
where 
\begin{equation}
(UUUU)_{\alpha\beta \: ;\: ij}
= U_{\alpha i}U^*_{\alpha j}U^*_{\beta i}U_{\beta j}
\end{equation}
etc.
We note that the last term in (\ref{eqn:deltaP1}) is at least second 
order in $a$. 

The energy eigenvalue $h_i$ can also be obtained to first order 
in $a$ as
\begin{equation}
h_i = \frac{m_i^2}{2E} + |U_{ei}|^2 a.
\end{equation}
Therefore, $I_{ij}(a)$ defined in (\ref{eqn:iij}) can be given by 
\begin{equation}
I_{ij}(a)=-\frac{\Delta m^2_{ij}}{2E}L + (|U_{ei}|^2-|U_{ej}|^2)
\int_0^L dx a(x), 
\end{equation}
where $\Delta m^2_{ij} \equiv m^2_j-m^2_i$. 
Up to this point one discussion relies only on the hierarchy 
$a \ll |\Delta m^2_{ij}| /E$ and the hierarchy among $\Delta m^2_{ij}$ 
need not to be assumed. 

To obtain an explicit form of $\Delta P$ we need the expression 
of $UUU\delta V$. Due to the mass hierarchy (\ref {eqn:assatm}) 
and (\ref{eqn:assdark}), we can combine the terms and calculate  
$\sum_{i=1,2}(UUU\delta V)_{\alpha, \beta \: ;\: i3}$ by ignoring 
the difference between $\Delta m^2_{13}$ and $\Delta m^2_{23}$. 
We also calculate the sum $(UUU\delta V)_{\alpha, \beta \: ;\: 12}$ 
separately: 
\begin{equation}
\sum_{i=1,2}(UUU\delta V)_{\alpha\beta \: ;\: i3} =
\frac{4Ea}{\Delta M^2}|U_{e 3}|^2 
\biggl[
2|U_{\alpha 3}|^2|U_{\beta 3}|^2 - (\delta_{\alpha e} |U_{\beta 3}|^2
+\delta_{\beta e}|U_{\alpha 3}|^2) 
\biggr]
\label{eqn:UUUDV1}
\end{equation}
\begin{eqnarray}
&&(UUU\delta V)_{\alpha\beta \: ;\: 12} \cr
&&=
-\frac{2Ea}{\Delta m^2}\biggl[U^*_{e1}U_{e2}U_{\alpha 1}U^*_{\alpha 2}
(|U_{\beta 2}|^2-|U_{\beta 1}|^2) 
+ U_{e 1}U^*_{e2}U^*_{\beta 1}U_{\beta 2}
(|U_{\alpha 2}|^2-|U_{\alpha 1}|^2)\biggr] \cr
&&-\frac{2Ea}{\Delta M^2}\biggl[U_{\alpha 1}U^*_{\alpha 2}
(U^*_{\beta 1}U_{\beta 3}U_{e2}U^*_{e3} 
+ U_{\beta 2}U^*_{\beta 3}U^*_{e1}U_{e3})
+ U^*_{\beta 1}U_{\beta 2}(U_{\alpha 1}U^*_{\alpha 3}U^*_{e2}U_{e3} 
+ U^*_{\alpha 2}U_{\alpha 3}U_{e1}U^*_{e3}) \biggr].
\label{eqn:UUUDV2}
\end{eqnarray}
In these equations we have dropped the terms further down by 
$\frac {\Delta m^2}{\Delta M^2}$. 
Notice that the terms containing $\Delta m^2$ cancel out in
$\sum_{i=1,2}(UUU\delta V)_{\alpha\beta \: ;\: i3}$. Note also
that only the mixing matrix elements $U_{\alpha 3}$ appear in
(\ref{eqn:UUUDV1}) as it occurs in the oscillation probability
in vacuum. 

Combining all these together we obtain the expression of 
$\Delta P$ which is valid to first order in $a$: 
\begin{eqnarray}
&&\Delta P(\nu_{\beta}\rightarrow\nu_{\alpha})\equiv 
P(\nu_\beta \to \nu_\alpha)-P(\bar{\nu_\beta}\to \bar{\nu_\alpha})\cr
&& =
-4 J 
\cos\biggl[\biggl(|U_{e1}|^2- |U_{e2}|^2\biggr)\int_0^L dx a(x)\biggr] 
\sin\biggl(\frac{\Delta m^2}{2E} L\biggr) \cr
&& 
+ 4 \mbox{Re}(U_{\alpha 1}U_{\alpha 2}^*U_{\beta 1}^*U_{\beta 2} )
\sin\biggl[\biggl(|U_{e1}|^2- |U_{e2}|^2\biggr) \int_0^L dx a(x)\biggr] 
\sin\biggl(\frac{\Delta m^2}{2E} L\biggr) \cr
&& + \frac{16Ea}{\Delta m^2}
\mbox{Re}\biggl[U_{\alpha 1}U_{\alpha 2}^* U^*_{e 1}U_{e 2}
\biggl(|U_{\beta 2}|^2- |U_{\beta 1}|^2 \biggr)  \cr
&& + U_{\beta 1}^* U_{\beta 2} U_{e 1}U^*_{e 2}
  \biggl(|U_{\alpha 2}|^2- |U_{\alpha 1}|^2 \biggr) \biggr]
\sin^2\biggl(\frac{\Delta m^2}{4E} L\biggr) \cr
&& +\frac{16Ea}{\Delta M^2}
\mbox{Re}\biggl[U_{\alpha 1}U^*_{\alpha 2}
(U^*_{\beta 1}U_{\beta 3}U_{e2}U^*_{e3}
+ U_{\beta 2}U^*_{\beta 3}U^*_{e1}U_{e3}) \cr
&& + U^*_{\beta 1}U_{\beta 2}(U_{\alpha 1}U^*_{\alpha 3}U^*_{e2}U_{e3}
+ U^*_{\alpha 2}U_{\alpha 3}U_{e1}U^*_{e3}) \biggr]
\sin^2\biggl(\frac{\Delta m^2}{4E}L\biggr) \cr
&& -\frac{32Ea}{\Delta M^2}
|U_{e 3}|^2 \biggl[
2|U_{\alpha 3}|^2|U_{\beta 3}|^2 - \biggl(\delta_{\alpha e}|U_{\beta 3}|^2 
+ \delta_{\beta e}|U_{\alpha 3}|^2\biggr) \biggr]  
\sin^2\biggl(\frac{\Delta M^2}{4E}L\biggr)
\label{eqn:deltaP2}
\end{eqnarray}
where the term proportional to $\mbox{Im}(UUU\delta V)$ and the 
$a$-dependent piece in the last term in (\ref{eqn:deltaP2}) are 
ignored because it is of order $a^2$ or higher. 

A few remarks are in order concerning the sign of $\Delta M^2$ 
and $\Delta m^2$. Strictly speaking, the matter effect distinguishes 
between the neutrino mass spectra of the types (A-1) and (A-2), or 
(B-1) and (B-2) defined in Fig. 2. 
However, with our choice of mass hierarchy, 
$\Delta P(\nu_{\beta}\rightarrow\nu_{\alpha})$
barely depends on $\Delta M^2$, and consequently the final results do 
not depend on the sign of $\Delta M^2$. It is because the last two 
terms in (\ref{eqn:deltaP2}) can be safely neglected due to the 
hierarchy in energy scales, 
\begin{equation} 
\frac{Ea}{\Delta M^2} = 1.04 \times 10^{-4}
\left(\frac{\rho}{2.72\mbox{gcm}^{-3}}\right)
\left(\frac{E}{1\mbox{GeV}}\right)
\left(\frac{\Delta M^2}{1\mbox{eV}^2}\right)^{-1}.
\end{equation} 
Hence, we do not distinguish between the mass patterns (A-1) and (A-2),  
or (B-1) and (B-2) in this work. (Note, however, that they can be 
distinguished by consideration of $r$-process nucleosynthesis in 
supernova \cite {Qian}.) 
For the smaller mass difference, $\Delta m^2 \equiv m_2^2-m_1^2$, 
we assume that it is positive. But, it is easy to accommodate the 
case with negative $\Delta m^2$ in our analysis after ignoring the 
last two terms in (\ref{eqn:deltaP2}), which we will do in our 
subsequent analysis. All we have to do is to change the over-all 
sign of $\Delta P(\nu_{\beta}\rightarrow\nu_{\alpha})$. 

By using the explicit form of the mixing matrix (\ref{eqn:ckmmatrix})
it is then straightforward to obtain $\Delta P$, the neutrino-antineutrino 
difference between oscillation probabilities, for 
$\nu_\mu \rightarrow \nu_e$ and 
$\nu_\mu \rightarrow \nu_\tau$ channels: 

\begin{eqnarray}
&&
\Delta P(\nu_\mu \to \nu_e) \equiv
P(\nu_\mu \to \nu_e)-P(\bar{\nu_\mu}\to \bar{\nu_e})\cr
&& 
= -4J \cos\biggl[c^2_{13}\cos 2\theta_{12}\int_0^L dxa(x)\biggr]
\sin\biggl(\frac{\Delta m^2L}{2E}\biggr) \cr
&& 
+ \biggl[ -\sin^2 2\theta_{12}c^2_{13}(c^2_{23}-s^2_{23}s^2_{13})
-4\cos 2\theta_{12} J\cot\delta \biggr]
\sin\biggl[c^2_{13}\cos 2\theta_{12}\int_0^L dxa(x) \biggr] 
\sin \biggl(\frac{\Delta m^2 L}{2E} \biggr) \cr
&& +4 \frac{Ea}{\Delta m^2} c^2_{13} 
\biggl[\sin 2\theta_{12}\sin 4\theta_{12}c^2_{13}(c^2_{23}-s^2_{23}s^2_{13}) 
+ 4 \cos 4\theta_{12} J \cot\delta \biggr]
\sin^2\biggl(\frac{\Delta m^2L}{4E}\biggr) 
\label{eqn:deltaPmue} 
\end{eqnarray}
\begin{eqnarray}
&&\Delta P(\nu_\mu \to \nu_\tau) \equiv 
P(\nu_\mu \to \nu_\tau)-P(\bar{\nu_\mu}\to \bar{\nu_\tau})\cr
&& 
=4J \cos\biggl[c^2_{13}\cos2\theta_{12}\int_0^L dxa(x)\biggr]
\sin \biggl(\frac{\Delta m^2 L}{2E} \biggr) \cr
&& 
+ \biggl[\{\biggl(\frac{1+s^2_{13}}{2}\biggr)
\cos2\theta_{12}\cos2\theta_{23}
-4c_{13}^{-2}J\cot\delta \}^2 \cr
&& 
+ \frac{1}{4}c^4_{13}(\sin^2 2\theta_{12}+\sin^2 2\theta_{23})
-\biggl(\frac{1+s^2_{13}}{2}\biggr)^2 \biggr]
\sin\biggl[c^2_{13}\cos 2\theta_{12}\int_0^L dxa(x) \biggr] 
\sin \biggl(\frac{\Delta m^2 L}{2E} \biggr) \cr
&& -4\frac{Ea}{\Delta m^2} \biggl[ 
\sin2\theta_{12}c^2_{13} \{\sin4\theta_{12}\sin^2 2\theta_{23}
\biggl(\frac{1+s^2_{13}}{2}\biggr)^2 - \sin4\theta_{12}s^2_{13}\} \cr
&& -4\cos4\theta_{12}\cos2\theta_{23} 
\biggl(\frac {1+s^2_{13}}{2}\biggr)J\cot \delta 
+ 32\cos2\theta_{12}c^{-2}_{13}J^2\cot^2\delta \biggr]
\sin^2\biggl(\frac{\Delta m^2L}{4E}\biggr) 
\label{eqn:deltaPmutau}
\end{eqnarray}
For completeness we also give the expression of $\Delta P$ for 
$\nu_e \rightarrow \nu_\tau$ channel:
\begin{eqnarray}
&&\Delta P(\nu_e\rightarrow\nu_\tau) \equiv P(\nu_e\rightarrow\nu_\tau) 
- P(\bar{\nu_e}\rightarrow\bar{\nu_\tau}) \cr
&& =
-4J \cos\biggl[c^2_{13}\cos 2\theta_{12}\int_0^L dxa(x)\biggr] 
\sin \biggl(\frac{\Delta m^2 L}{2E} \biggr) \cr 
&&
+ \biggl[ -\sin^2 2\theta_{12}c^2_{13}(s^2_{23}-c^2_{23}s^2_{13})
+4\cos 2\theta_{12}J\cot\delta \biggr] 
\sin \biggl[c^2_{13}\cos 2\theta_{12}\int_0^L dxa(x)\biggr] 
\sin \biggl(\frac{\Delta m^2 L}{2E} \biggr) \cr
&& +4\frac{Ea}{\Delta m^2}c^2_{13} 
\biggl[\sin2\theta_{12}\sin4\theta_{12}c^2_{13}
(s^2_{23}-c^2_{23}s^2_{13})-4\cos4\theta_{12} J \cot\delta \biggr] 
\sin^2\biggl(\frac{\Delta m^2L}{4E}\biggr) 
\label{eqn:deltaPetau}
\end{eqnarray}

Let us compare these analytic results with the exact solutions 
obtained in \cite {Zaglauer} for a constant matter density to 
examine the accuracy of our approximate formulas.
We pick up the following two parameter sets,
\begin{eqnarray}
&\mbox{(a)}& \ \ s_{23}^2  = 3.0\times 10^{-3}, s_{13}^2 = 2.0\times 10^{-2}\\
\label{eqn:seta}
&\mbox{(b)}&\ \ s_{23}^2 = 2.0\times 10^{-2}, s_{13}^2 = 0.98
\label{eqn:setb}
\end{eqnarray}
from the allowed regions (A) and (B), respectively. These parameters 
are chosen so that $J$ takes a maximal value within the each allowed 
region and are plotted in Fig. 1. We will use the same parameter sets 
also in our analyses in the following sections. To obtain the ``exact 
results'' by using the analytic expressions 
given in \cite{Zaglauer} we take an average over the rapid 
oscillations due to the larger mass difference $\Delta M^2$. We take 
the constant matter density and electron fraction, 
$\rho$ = 2.72 {g/cm}$^3$ and $Y_e=0.5$. 
In Fig. 3 we show the comparison between the exact and 
approximated values of $\Delta P(\nu_\mu\to\nu_e)$ and 
$\Delta P(\nu_\mu\to\nu_\tau)$.  We have fixed the remaining 
parameters as $s_{12}^2 =0.3$, $\Delta M^2 = 5$ eV$^2$ and 
$\delta = \pi/2$. 
We see that for $E\sim$ 1 GeV the approximation is very good even 
for smaller values of $\Delta m^2$ ($\sim 10^{-3} $eV$^2$ ) than the 
one we assumed in eq. (\ref{eqn:assatm}). 

%
\section{CP Violation vs. Matter Effect in Long-Baseline Neutrino
Oscillation Experiments: an order-of-magnitude analysis}

In this section we present the results of our order estimations of 
the magnitude of $CP$ violation, and give an answer to the question 
of relative importance between the genuine $CP$-violating 
and the matter effects. This precedes the presentation of the results 
of detailed numerical computation in the next section, which will 
provide us complementary informations. The order-of-magnitude 
estimation based on our analytic expressions of $\Delta P$ illuminates 
the general features of the interplay between the $CP$ phase and the 
matter effects, and is valid in whole allowed parameter regions. 
On the other hand, the numerical computation using the exact 
expression of the oscillation probability (albeit under the 
constant density ansatz) will reveal precise features of the relative 
importance between competing two effects. 

For convenience of our discussion let us denote the three terms in 
(\ref{eqn:deltaPmue}) and (\ref{eqn:deltaPmutau}) as 
\begin{eqnarray}\nonumber
&&\Delta P(\nu_\beta\rightarrow\nu_\alpha) = \Delta P_{\beta \alpha}^{CP} + 
\Delta P_{\beta \alpha}^{matter 1}+
\Delta P_{\beta \alpha}^{matter 2} \\
&&\equiv 
C_{\beta \alpha}^{CP}\sin\Delta  + 
C_{\beta \alpha}^{matter1}\sin\Delta +
C_{\beta \alpha}^{matter2}\sin^2(\Delta/2), 
\ \  (\beta= \mu,\ \alpha=e, \tau\ ),\ \ 
\end{eqnarray}
where $\Delta \equiv \Delta m^2 L/2E$, which can be 
of order 1 for $\Delta m^2 \lsim 10^{-2}$ eV$^2$, 
$E \gsim 1$ GeV and $L \gsim$ 100 km. 
$C_{\beta \alpha}^{CP}$ is the coefficient of the first term 
in these equations which represents the genuine $CP$ violating 
effect corrected by the matter effect. 
$C_{\beta \alpha}^{matter1}$ and $C_{\beta \alpha}^{matter2}$ are 
the coefficients in the second and third terms in 
(\ref{eqn:deltaPmue}) and (\ref{eqn:deltaPmutau}), display the 
matter effects from the different sources. That is, 
$C_{\beta \alpha}^{matter1}$ represents the matter effect 
which arises due to the evolution of the phase of the neutrino wave 
function in matter, whereas $C_{\beta \alpha}^{matter2}$ 
comes from the correction to the mixing matrix, as exhibited in 
the $UUU\delta V$ factors in (\ref {eqn:deltaP1}).
We notice that
$C_{\beta \alpha}^{CP}$ and 
$C_{\beta \alpha}^{matter1}$ contain the integral 
$\int_0^L dx a(x)$ which can be estimated, by assuming the 
constant density 2.72 g/cm$^3$, as 
$aL=1.2 \times 10^{-1}$ for KEK-PS$\rightarrow$Super-Kamiokande 
experiment with baseline of 250 km, and 
$aL=3.5 \times 10^{-1}$ for MINOS or CERN-ICARUS
experiment with baseline of 730 km or 732 km, respectively. 
On the other hand, $C_{\beta \alpha}^{matter2}$ carries 
the coefficient $Ea/\Delta m^2$ which is of the order of 
$\sim 10^{-2}$ for $E\sim 1$ GeV and $\Delta m^2 \sim 10^{-2}$ eV$^2$. 
Therefore, apart from prefactors which depend on mixing angles, we 
estimate that  
\begin{eqnarray}
C_{\beta \alpha}^{CP} &\propto& -4J \sim 10^{-2},
\\
C_{\beta \alpha}^{matter1} &\propto& aL \sim 0.1,\\ 
C_{\beta \alpha}^{matter2} &\propto& \frac{2Ea}{\Delta m^2}
\sim 10^{-2}.
\label{eqn:matter2} 
\end{eqnarray} 
We recognize that $C_{\beta \alpha}^{matter1}$ can be much 
larger, depending on the mixing angles, than the genuine 
$CP$ violating effect $C_{\beta \alpha}^{CP}$. 
We also note that from eq. (\ref{eqn:matter2}) it is clear 
that if we assume smaller $\Delta m^2$ values 
$C_{\beta \alpha}^{matter2}$ 
become larger (See also the Tables 1 and 2). 

In Table 1 we summarize the results of our order-of-magnitude 
estimation of these three coefficients for $\nu_\mu \rightarrow \nu_e$
channel. In doing the estimation we have taken into account the 
coefficient which depend on the mixing angles, and the numbers 
presented in Table 1 refers to the possible maximal values in 
each region. 
We notice that, if $\sin \Delta \sim 1$,  in the region (A) 
the oscillation probability 
$P(\nu_\mu \to \nu_e)$ can be large, $\sim 1$, but $CP$ violation 
due to the matter effect is much larger than the intrinsic $CP$ 
violation effect,  
$\Delta P^{CP} \sim 0.1\times \Delta P^{matter}$.  
On the other hand, in the region (B), $P(\nu_\mu \to \nu_e)$ is 
small, $\sim 10^{-2}$, but the contamination of matter effect 
in $CP$ violation is very small, $\Delta P^{matter} \sim 10^{-4}$ 
compared to $\Delta P^{CP}\sim 10^{-2}$. 

In Table 2, we present the same quantities for 
$\nu_\mu \rightarrow \nu_\tau$ channel.
We observe that in this channel the intrinsic $CP$ violating 
effect is larger than the matter effect in both regions 
(A) and (B), i.e., $\Delta P^{CP}\sim 10^{-2}$
whereas $\Delta P^{matter}\sim 10^{-3}$.

Let us try to understand the qualitative features of these results. 
Probably, the most interesting aspect of our results is that the 
matter effect is small in the region (B) in 
$\nu_\mu \rightarrow \nu_e$ channel. But, in fact it is not difficult 
to understand the reason why. In the region (B) the mixing
parameters are such that $\nu_e$ is much heavier than
$\nu_\mu$ and $\nu_\tau$, which are almost degenerate, or that
$\nu_\mu$ and $\nu_\tau$ are heavier than $\nu_e$ by the same
amount in the mass hierarchies given in Fig. 2. 
The matter effect only affects electron neutrinos
and the matter potential is small compared with $\nu_e$-$\nu_\mu$
mass difference, $\Delta M^2/E \gg a$. Then, it is easy to expect
that the matter effect is small not only in
$\nu_\mu \rightarrow \nu_\tau$ but also in
$\nu_\mu \rightarrow \nu_e$ channels.

In the region (A), on the other hand, $\nu_\tau$ is much heavier 
than $\nu_e$ and $\nu_\mu$, or vice versa, and $\nu_e$ and $\nu_\mu$ 
are strongly mixed. Therefore, one would naively expect that the matter
effect is sizable, as it is indeed the case in 
$\nu_\mu \rightarrow \nu_e$ channel.
But, the situation is different in $\nu_\mu \rightarrow \nu_\tau$ 
channel. A $\nu_\mu$ can easily communicate with $\nu_e$ and thus 
feels the effect of earth matter but to make oscillation into 
$\nu_\tau$ it has to overcome the huge mass difference compared 
with the matter potential. Therefore, the matter effect is not 
dominant in the region (A) of $\nu_\mu \rightarrow \nu_\tau$ channel.

\section{CP Violation vs. Matter Effect in Long-Baseline Neutrino
Oscillation Experiments: a numerical analysis}

In this section we present the results of our numerical analysis 
using the two sets of parameters (a) and (b) given in 
(\ref{eqn:seta}) and (\ref{eqn:setb}).  
We do this first for $\nu_\mu \to \nu_e$ channel, and second for 
$\nu_\mu \to \nu_\tau$ channel. 
All the calculations are carried out by using the exact analytic 
expressions found in \cite{Zaglauer} with the procedure mentioned 
at the end of section V. 

\subsection{$\bf \nu_\mu \to \nu_e$ and 
$\bar{\nu}_\mu\to\bar{\nu}_e$ channels}

First let us take the parameter set (a). 
In Fig. 4 we plot 
$P(\nu_\mu\to\nu_e)$ and 
$P(\bar{\nu}_\mu\to\bar{\nu}_e)$ 
and the corresponding $\Delta P(\nu_\mu\to\nu_e)\times 100$ 
as a function of $\Delta m^2/E$ for different values of $s_{12}^2$. 
We fix the distance $L$ = 250 and 730 km in Figs. 4(a) and 4(b), 
respectively. We see that from (iv) and (vi) in Fig. 4 that 
$\Delta P(\nu_\mu\to\nu_e)$ can be as large as $\sim$ 10 \% for 
$L=250$ km and $\sim$ 25 \% for 
$L=730$ km due to the the matter effect whereas the intrinsic $CP$ 
violation is at most $\sim 1.5$ \%, in agreement with our estimation in the 
previous section (see Table 1). 
We can conclude that the matter effect dominates over genuine $CP$ 
violating effect except at the exceptional point $s_{12}^2 \simeq$ 0.5. 
As we can see from the second term in eq. (\ref{eqn:deltaPmue}) that 
the matter potential is multiplied by $\cos2\theta_{12}$ and hence 
the matter effect in $\Delta P(\nu_\mu\to\nu_e)$ is suppressed if 
$s_{12}^2$ is close to 0.5. This suppression can also be seen in 
Fig. 3 in the first reference in \cite{Tanimoto}. We also notice that 
this factor $\cos2\theta_{12}$ gives rise to the sign difference in 
$\Delta P(\nu_\mu\to\nu_e)$ for $s_{12}^2 >0.5$ 
and $s_{12}^2 <0.5$ as we can confirm from (iv) and (vi) in Fig. 4. 

In  Fig. 5 we plot the contour of $\Delta P(\nu_\mu\to\nu_e)$ on the 
$s_{12}^2-\Delta m^2/E$ plane to see the global features of the 
coexisting $CP$ and the matter effects. We see that the contours 
for the pure $CP$ and the pure matter cases are very different 
and the amplitude of the latter is larger. It may be difficult to 
see the genuine $CP$ violating effect just by observing 
$\Delta P(\nu_\mu\to\nu_e)$ apart from the exceptional region 
$s_{12}^2\simeq 0.5$.  

Let us turn to the parameter set (b). In Fig. 6 we plot 
as in Fig. 4 $P(\nu_\mu\to\nu_e)$ (and for $\bar{\nu}$) and 
$\Delta P(\nu_\mu\to\nu_e)$ but for the parameter set (b). 
In this case the probabilities of $\nu_\mu\to\nu_e$ and 
$\bar{\nu}_\mu\to\bar{\nu}_e$ are small, of the order of 1 \%, as we 
can see in Fig. 6 and they may be denoted as the minor channels. In 
this case the matter effect contamination in $\Delta P$ is negligibly 
small, in agreement with our estimation done in the previous section.  
We confirm that the genuine $CP$ violating effect dominates 
over the matter effect at the parameter (b). 

\subsection{\bf $\nu_\mu \to \nu_\tau$ and $\bar{\nu}_\mu\to\bar{\nu}_\tau$ 
channels}

In Fig. 7 we plot $P$ and $\Delta P$ as in Fig. 4 but for the parameter 
set (a) in the $\nu_\mu\to\nu_\tau$ and $\bar{\nu}_\mu\to\bar{\nu}_\tau$ 
channels. These channels are also minor since $P \lsim 2$ \% but 
$\Delta P_{\mu \tau}$ is relatively large  $\sim 1$ \%. 
We see that the genuine $CP$ violating effect is larger than 
the matter effect. In Fig. 8 we plot as in Fig. 5 the contour of 
$\Delta P(\nu_\mu\to\nu_\tau)$ on the $s_{12}^2-\Delta m^2/E$ plane. 
We confirm from these contours that the genuine $CP$ effect is larger 
than the matter effect, in agreement with our estimation given in Table 2. 

In Fig. 9 we plot $P$ and $\Delta P$ for the parameter set (b). 
In this case $\nu_\mu \to \nu_\tau$ is the dominant channel but 
$\Delta P$ is small $\sim 0.7$ \%. 
We see that apart from the small $s_{12}^2$ region for the baseline 
$L=730$ km the genuine $CP$ violating effect is larger than the 
matter effect. Hence, we conclude that for $\nu_\mu \to \nu_\tau$ and 
$\bar{\nu}_\mu\to\bar{\nu}_\tau$ channels are relatively 
free from the matter effect. 
In Fig. 10 we plot the contour of $\Delta P(\nu_\mu\to\nu_\tau)$
for this parameter set. We see that also for this case 
the  genuine $CP$ effect is larger than the matter effect in agreement 
with our prediction (see Table 2).

\section{Conclusion}

We have investigated in detail the $CP$ violation in 
long-baseline neutrino oscillation experiments in the 
presence of matter effect under the assumption of neutrino mass 
hierarchy, $\Delta m^2\sim 10^{-2}$ eV$^2$ and
$\Delta M^2\sim $ a few eV$^2$, motivated by the atmospheric 
neutrino anomaly and the cosmological dark matter. 
We developed the matter perturbation theory using the hierarchy 
in energy scales, $a \ll \frac{\Delta m^2}{2E}$, and derived the 
approximate analytic expressions of $\nu-\bar{\nu}$ difference
in oscillation probabilities $\Delta P(\nu_\beta\to\nu_\alpha)$. 
We have found that in a good approximation $\Delta P$ can be expressed 
as a sum of three terms which represent the genuine $CP$ violating 
effect and two different correction terms due to the matter effect.  
These analytic expressions of the $\Delta P$ are useful in 
understanding the global features of the competition of two effects. 

We have studied the question of how large is the magnitude of 
$CP$ violations due to intrinsic $CP$-violating phase and to the 
matter effect in the earth. The assumed mass hierarchy mentioned 
above and the interpretation of the atmospheric neutrino anomaly 
in terms of the neutrino oscillations allow us to restrict ourselves 
into the parameter regions strongly constrained by the reactor and 
the accelerator experiments, (A) small-$s_{13}$ and small-$s_{23}$
and (B) large-$s_{13}$ and arbitrary $s_{23}$. 

We have found the following structure 
(as summarized in Tables 1 and 2). 

$\nu_\mu\to\nu_e$ channel: 
In the region (A) the matter effect contamination is much larger than 
the genuine $CP$ violation effect except for the case
$s_{12}\simeq0.5$.  Whereas in (B) the the matter effect contamination 
is negligible compared with genuine $CP$ violation. Unfortunately, 
the magnitude of the latter is small, $\sim 0.5$ \%. 

$\nu_\mu\to\nu_\tau$ channel:  
The matter contamination is smaller than the genuine $CP$ violation 
effect and is not harmful in both regions (A) and (B). The magnitude 
of the $CP$ violation is again not sizable, $\sim 1$ \%.

Thus, if the mass hierarchy motivated by the dark matter is the truth 
in nature, it appears to be necessary to invent new method for 
measuring $CP$ violation of $\sim 1$ \% level in long-baseline 
neutrino oscillation experiments. We are planning to discuss 
an idea toward the goal elsewhere \cite {MN2}.

\vglue 1cm
\noindent
Note added: While we were to complete this paper we became aware 
of the paper by Arafune et al. \cite {AKS} which addresses the 
similar topics. However, they consider different neutrino 
mass spectrum and employ a different approximation scheme from 
ours which requires that 
$\frac {aL}{E} \ll 1$ and $\frac {\Delta m^2 L}{E} \ll 1$, 
whereas we only need $\frac {Ea}{\Delta m^2} \ll 1$. 

We also note that after this paper had been submitted for 
publication, a new data on atmospheric neutrinos 
from Super-Kamiokande experiment has appeared \cite{superkam}. 
The data seem to favor smaller values of $\Delta m^2$ than 
the one we assumed in this paper. However, the results of the 
analysis done in this paper is presented so that they are useful 
for such smaller values of $\Delta m^2$. 
We stress that the perturbative treatment of the matter effect is 
still valid even for $\Delta m^2$ as small as $10^{-3}$ eV$^2$.

\begin{center}
{\large Acknowledgements}
\end{center}
We thank Masafumi Koike, Joe Sato and Osamu Yasuda for pointing out 
an error in our earlier version of this paper and for discussions 
which led us to the clarification of the point. 
One of us (H.M.) was supported partially by Grant-in-Aid for 
Scientific Research on Priority Areas under \#08237214 and is 
supported in part by Grant-in-Aid for Scientific Research \#09045036 
under International Scientific Research Program, Inter-University 
Cooperative Research.
The other (H.N.) has been supported by a DGICYT postdoctoral fellowship 
at Universitat de Val\`encia under Grant PB95-1077 
and TMR network ERBFMRXCT960090 of the European Union.

\newpage

\begin{center}
{\large APPENDIX}
\end{center}

We verify that the adiabatic approximation which we have employed
to obtain (\ref{eqn:solution}) is in fact a very good approximation
for the long-baseline neutrino experiments.
The adiabaticity condition is nothing but the condition
\begin{equation}
\left| \left[ \frac{d}{dx}V^+(x) \cdot V(x)\right]_{ij}\right|
\ll |(H_d)_{ii}|
\label{eqn:adiab}
\end{equation}
To first order in matter perturbation theory the off-diagonal elements
of the LHS of (\ref{eqn:adiab}) can be expressed as
\[
\left[
\frac{d}{dx}V^+(x) \cdot V(x) \right]_{ij} =
-a'(x)\frac{U_{ej}^*U_{ei}}{h_j^{(0)}-h_i^{(0)}}
\]
for $i\neq j$ and the diagonal elements vanish for $i=j$.
Then the adiabativicity condition can be written as
\begin{equation}
|U^*_{ej}U_{ei}| \ll \left|\frac{1}{a'(x)}\right|
\left(\frac{\Delta m^2}{2E}\right)^2
\label{eqn:adiabfinal}
\end{equation}
where the RHS is replaced by $\Delta m^2$, the smallest
$\Delta m^2_{ij}$. The expression (\ref{eqn:adiabfinal}) in fact 
represents the sufficient condition, but we will see that it is 
well satisfied.

The density gradient in the continental structure is at most
0.2g/$cm^3$ over the depth of 20 km \cite {Stacey}, which amount to
the baseline of 1000 km. Using this $a'(x)/a(x)$ can be estimated as
\[
\left| \frac{a'(x)}{a(x)}\right| \simeq 1.5 \times10^{-4}
\mbox{km}^{-1}
\]
Then, the the adiabaticity condition reads
\[
|U^*_{ej}U_{ei}| \ll 0.83 \times10^4
\left(\frac{\Delta m^2}{10^{-2}\mbox{eV}^2}
\right)^2\left(\frac{E}{1\mbox{GeV}}\right)^{-2}
\left(\frac{a'}{4\times10^{-4}\mbox{gcm}^{-3}\mbox{km}^{-1}}\right)^{-1}
\]
It is clear that the adiabaticity condition is well satisfied in the
long baseline experiments due to the slow variation of matter density
in the continental structure.


\newpage
\vglue 3.0cm
\begin{tabular}{|p{2.0cm}||p{2.5cm}|p{2.2cm}|p{2.2cm}|p{3.2cm}|}   
\hline
\ \ \ Region & \ $P(\nu_\mu \to \nu_e)$  
& \ \ $C_{\mu e}^{CP}$ & \ \ $C_{\mu e}^{matter1}$ 
& \ \ $C_{\mu e}^{matter2}$ \\ \hline\hline
\ \ \ \ (A)  &$ \  \  \sim 1$       &$\  \  \sim 10^{-2}  $ 
& $\  \  \sim 0.1$& $ \  \  \sim 10^{-2}\ (0.1) $   \\ \hline
\ \ \ \ (B)  &$\  \  \sim 10^{-2}$ &$\  \  \sim 10^{-2}$ 
& $\  \  \sim 10^{-4}$& $\  \  \sim 10^{-6}\ (10^{-5}) $   \\ \hline
\end{tabular}
\vglue 0.5cm
\noindent 
Table 1: The order of magnitude estimate of maximal values of 
$P$, $C^{CP}$ and $C^{matter1,2}$ for 
$\nu_\mu \to \nu_e$ channel in the parameter region 
(A) and (B) for $E \sim 1$ GeV and $\Delta m^2 \sim 10^{-2}$ eV$^{2}$. 
For $C_{\mu e}^{matter2}$ we also show in the parentheses 
the values for the case $\Delta m^2 \sim 10^{-3}$ eV$^{2}$.

\vglue 1.0cm
\begin{tabular}{|p{2.0cm}||p{2.5cm}|p{2.2cm}|p{2.2cm}|p{3.2cm}|}   
\hline
 \ \ \ Region & \ $P(\nu_\mu \to \nu_\tau)$  
& \ \ $C_{\mu \tau}^{CP}$
& \ \ $C_{\mu \tau}^{matter1}$ 
& \ \ $C_{\mu \tau}^{matter2}$
  \\ \hline\hline
\ \ \ \ (A)  &$\  \  \sim 10^{-2}$   &$\  \  \sim 10^{-2}$ 
& $\  \  \sim 10^{-3}$  & $\  \  \sim 10^{-4}\ (10^{-3})$  \\ \hline
\ \ \ \ (B)  &$\  \  \sim 1$ &$\  \  \sim 10^{-2}$ 
& $\  \  \sim 10^{-3}$ & $\  \  \sim 10^{-4} \ (10^{-3})$   \\ \hline
\end{tabular}
\vglue 0.5cm
\noindent 
Table 2: Same as Table 1 but for 
$\nu_\mu \to \nu_\tau$ channel. 
\vglue 1cm 

\newpage
\vglue 2.cm
\centerline{\hskip -1cm
\psfig{file=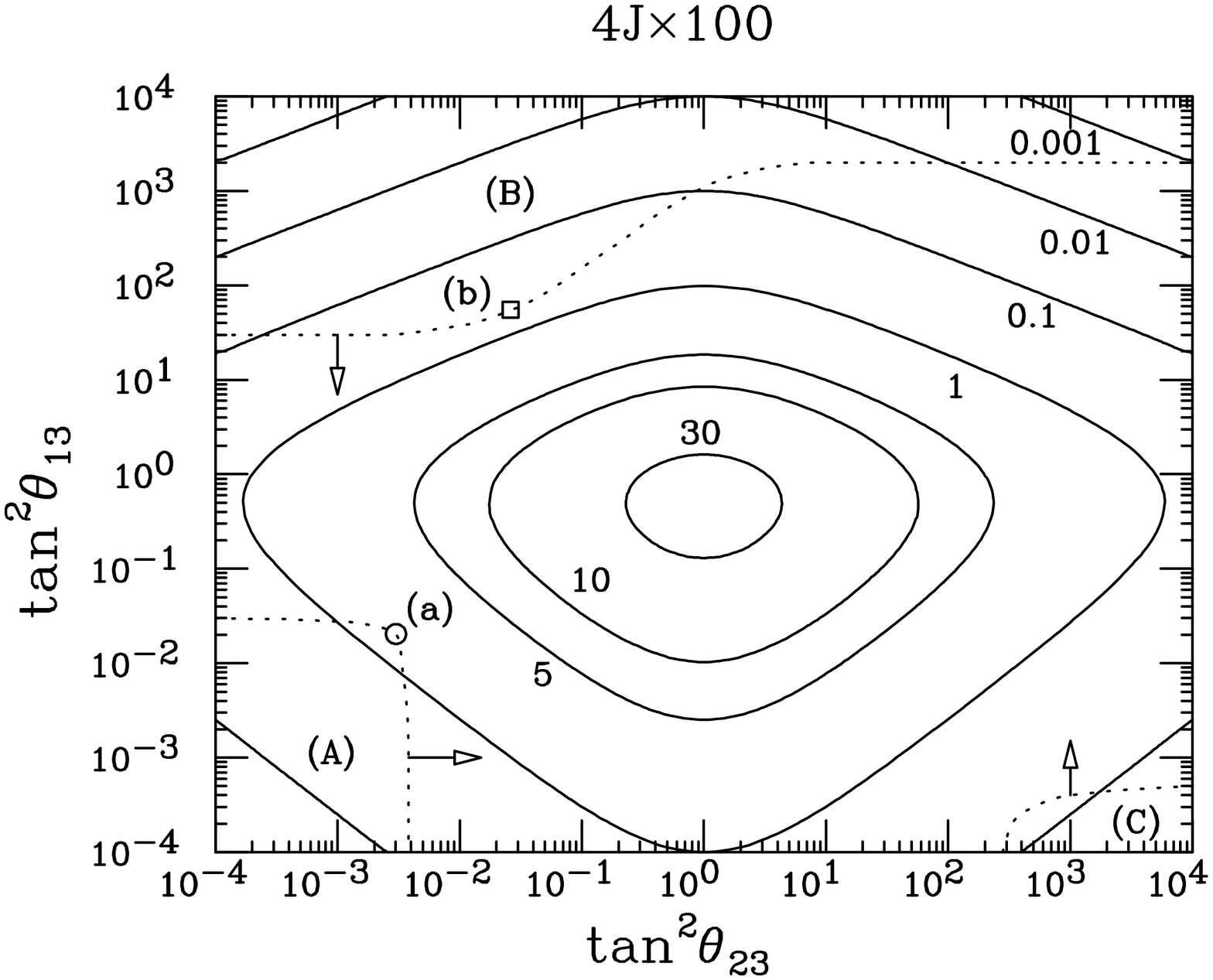,height=15.0cm,width=16.0cm,angle=90}}
\noindent
Fig. 1: Contour plot of 4$J\times$100 is shown where
$J \equiv c_{12}s_{12}c_{23}s_{23}c_{13}^2s_{13}\sin\delta$.
We set $\theta_{12} = \pi/4$ and $\delta = \pi/2$ which
maximize $J$.
The area indicated by dashed lines with arrows is the region
excluded by accelerator and reactor experiments at 90 \% C. L. 
obtained in ref. \cite{Fogli} for the case $\Delta M^2 = 5$ eV$^2$.
The points indicated by the (a) open circle and (b) square are
the parameters we will use in this work.

\newpage
\vglue 1.cm
\centerline{
\psfig{file=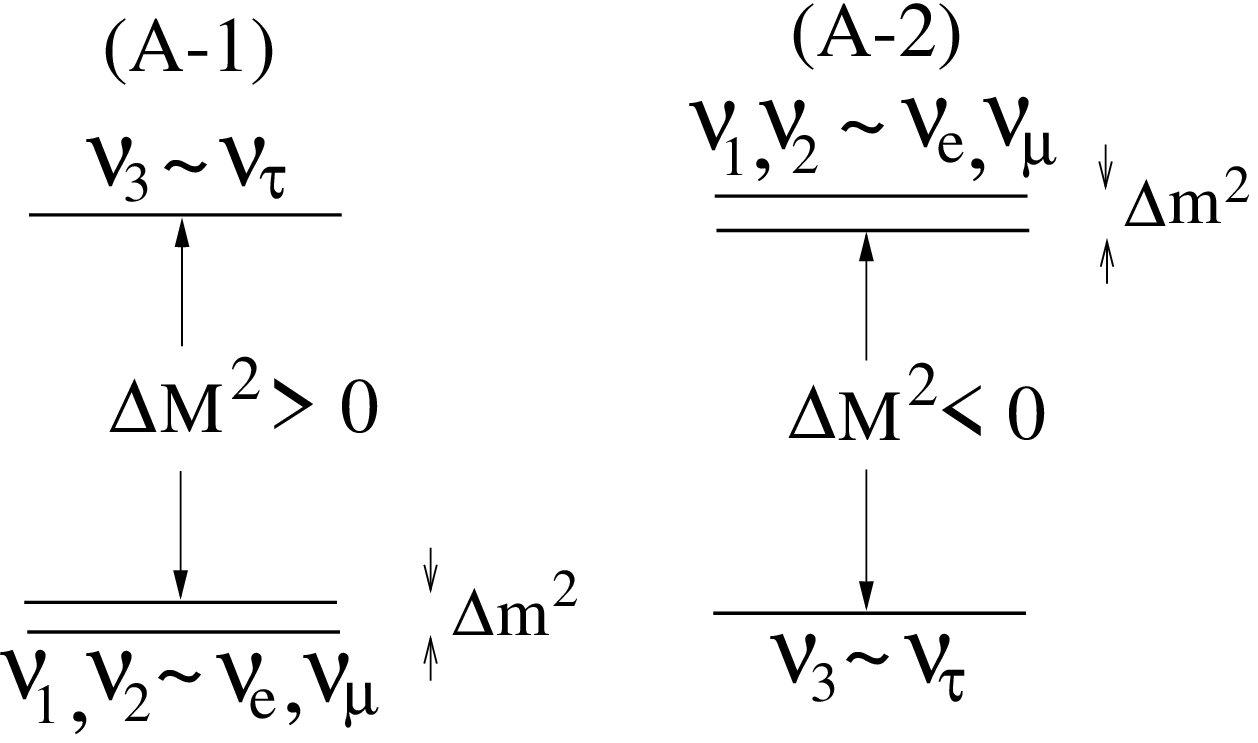,height=7.0cm,width=12.5cm,angle=-90}}
\vglue 1.5cm
\centerline{
\psfig{file=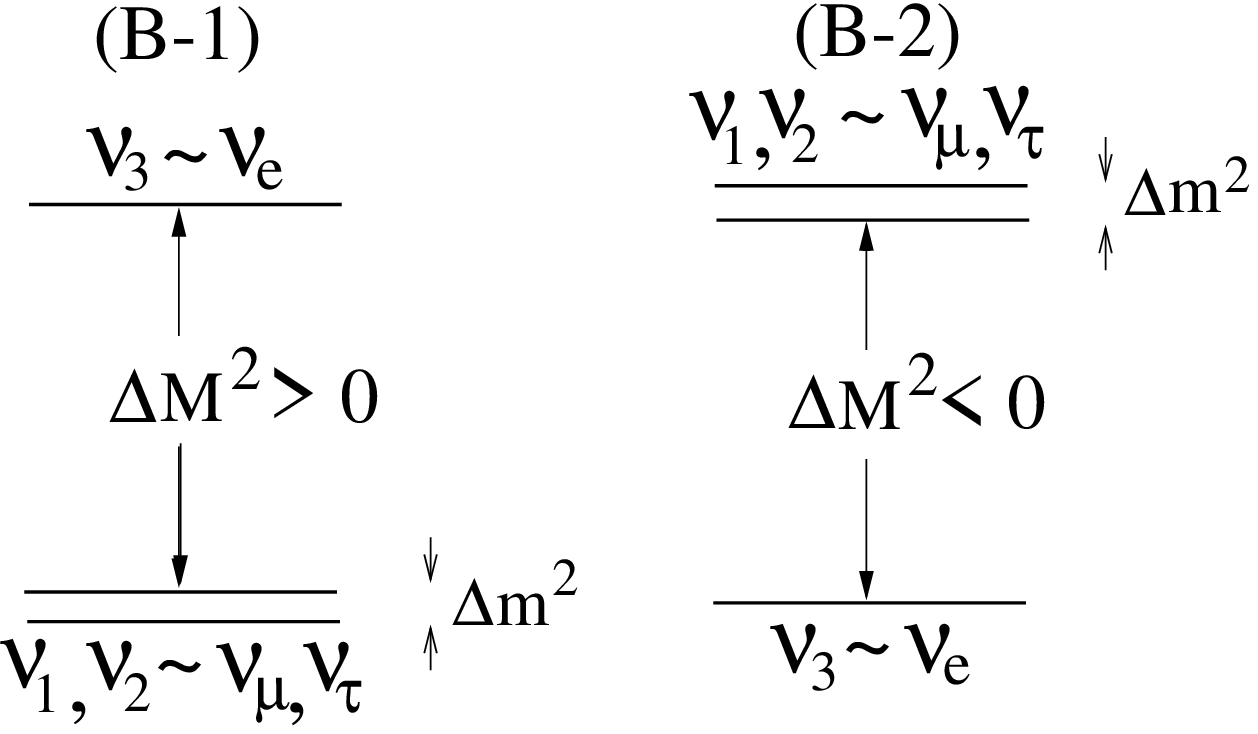,height=7.0cm,width=12.5cm,angle=-90}}
\vglue 0.8cm
\noindent
Fig. 2: The neutrino mass spectra which we consider in this paper. 
We assume that 
$\Delta M^2 \equiv m_3^2-m_1^2 \simeq m_3^2-m_2^2 \sim$ a few eV$^2$
and $\Delta m^2 \equiv m_2^2-m_1^2 \sim 10^{-2}$ eV$^2$.
The figure (A-1,2) and (B-1,2) are for the case where mixing parameters
are in the region (A) $s_{13}$,$s_{23}\ll 1$
and in the region (B) $s_{13}\sim 1$ and arbitrary $s_{23}$, respectively.
\newpage
\vglue 0.5cm
\centerline{\hskip 5.0cm
\psfig{file=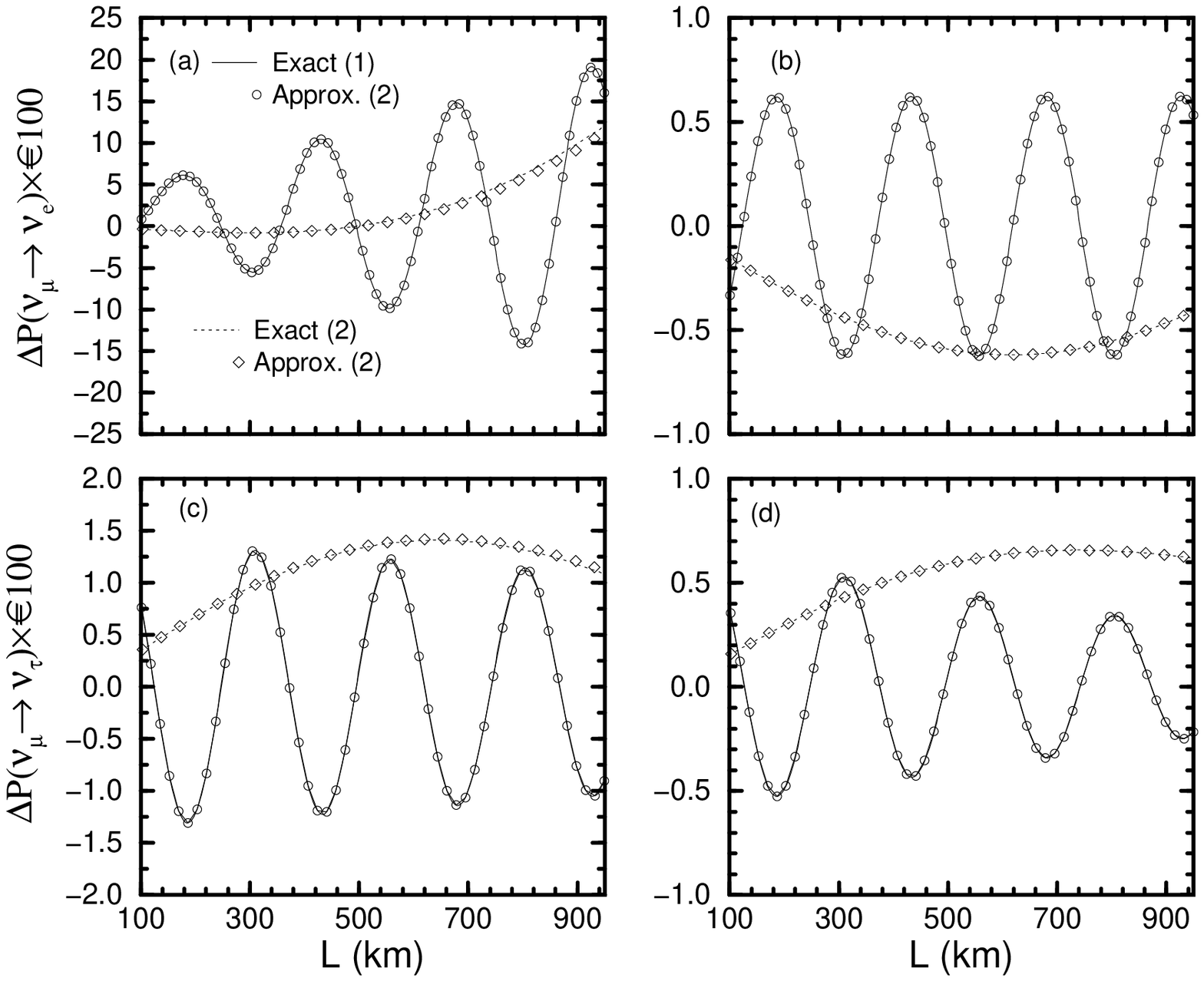,height=18.0cm,width=23.0cm,angle=-90}}
\vglue -5.0cm
\noindent
Fig. 3: We plot both the exact and approximated values of
$\Delta P(\nu_\mu\to\nu_e)$ (upper two panels) and 
$\Delta P(\nu_\mu\to\nu_\tau)$ (lower two panels) as a function
of distance L from the neutrino source.
We fixed the mixing parameters as $s_{23}^2 = 3.0\times 10^{-3}$,
$s_{13}^2 = 2.0\times 10^{-2}$ for the left two panels (a) and (c)
and $s_{23}^2 = 2.0\times 10^{-2}$, $s_{13}^2$ = 0.98 for
the right two panels (b) and (d). The remaining parameters are
fixed to be the same for all the case (a-d), i.e.,
$s_{12}^2 =0.3$, $\Delta M^2 = 5$ eV$^2$ and $\delta = \pi/2$.
The solid lines and open circles are for the case
$\Delta m^2/E = 10^{-2}$ eV$^2$/GeV and the dotted
and open diamonds are for the case
$\Delta m^2/E = 10^{-3}$ eV$^2$/ GeV.
\newpage
\vglue -1.5cm
\centerline{(a) $L$ = 250 km}
\centerline{\hskip -1cm
\psfig{file=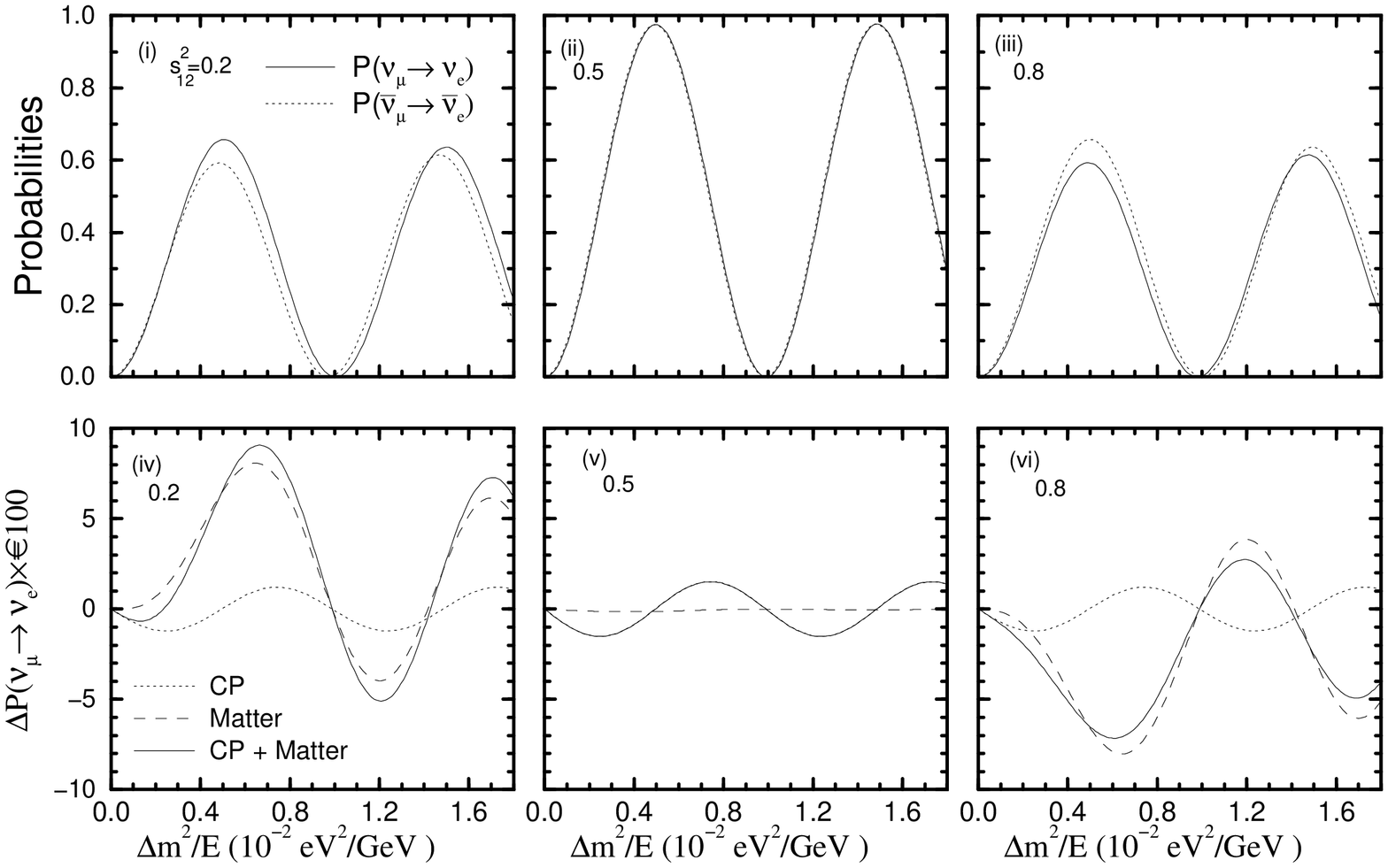,height=11.0cm,width=15.0cm,angle=-90}}
\vglue -3.0cm
\centerline{(b) $L$ = 730 km}
\centerline{\hskip -1cm
\psfig{file=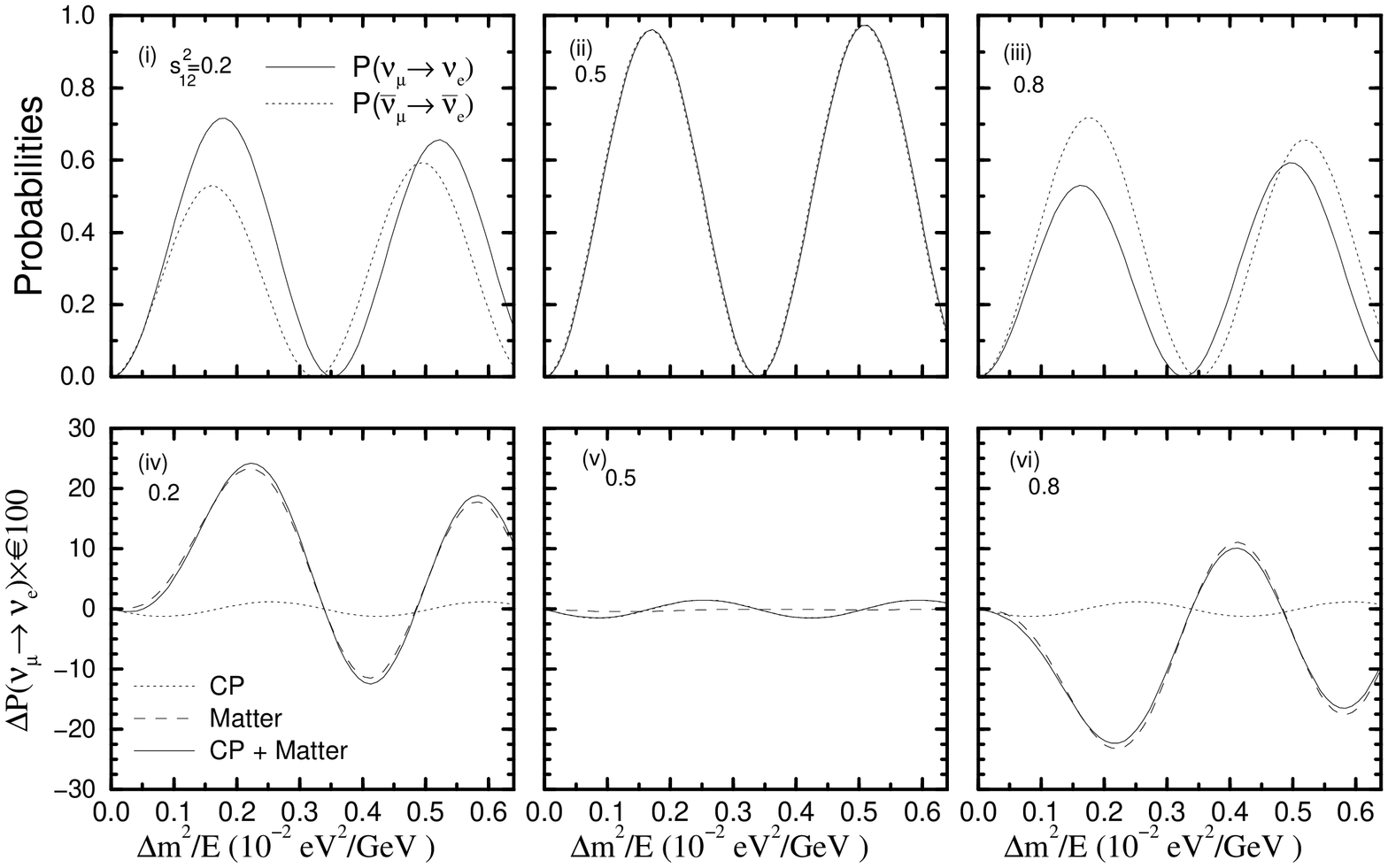,height=11.0cm,width=15.0cm,angle=-90}}
\vglue -3.0cm
\noindent
Fig. 4: We plot in (i), (ii) and (iii) $P(\nu_\mu\to\nu_e)$ and
$P(\bar{\nu}_\mu\to\bar{\nu}_e)$ for the parameter set
(a) $s_{23}^2 = 3.0\times 10^{-3}$, $s_{13}^2 = 2.0\times 10^{-2}$
as a function of $\Delta m^2/E$ for $s_{12}^2$ = 0.2, 0.5 and 0.8,
respectively.
We fixed the remaining parameters as $\Delta M^2 = 5$ eV$^2$,
$\delta = \pi/2$. The figures (a) and (b) are for $L$ = 250 km (a) and 
$L$ = 730 km (b), respectively. In (iv), (v) and (vi)
we plot the corresponding $\Delta P(\nu_\mu\to\nu_e)\times 100$
and the different curves are for the cases
where only the $CP$ (dotted line), only the matter (dashed line), and 
both the $CP$ and the matter (solid line) effects are taken into account.
\newpage
\vglue -1.5cm
\centerline{\hskip -1cm
\psfig{file=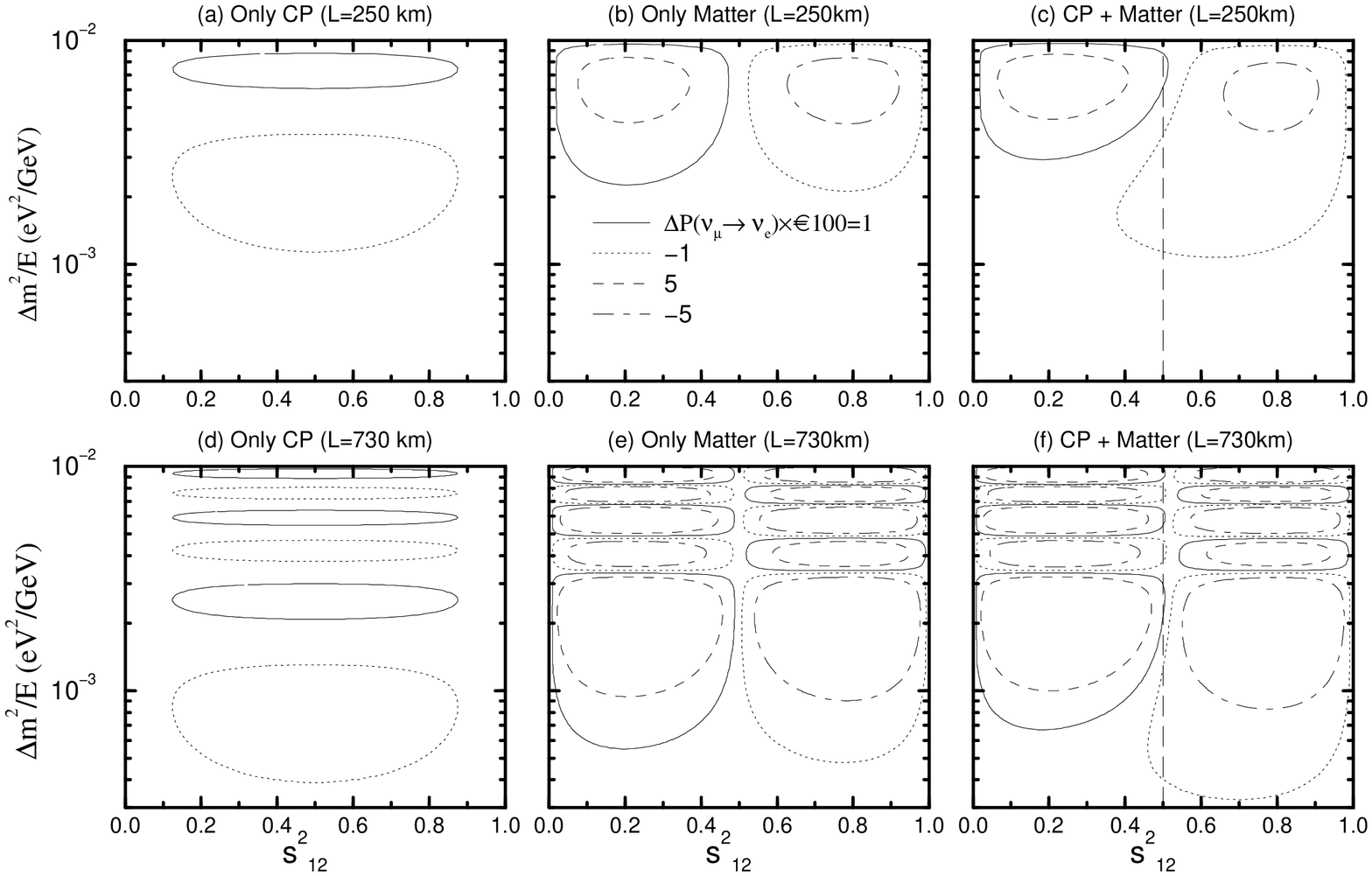,height=18.0cm,width=18.0cm,angle=-90}}
\vglue -1.5cm
\noindent
Fig. 5: We plot, for the parameter set 
(a) $s_{23}^2 = 3.0\times 10^{-3}$ and $s_{13}^2 = 2.0\times 10^{-2}$, 
the contour of $\Delta P(\nu_\mu\to\nu_e)\times 100$
in the $s_{12}^2-\Delta m^2/E$ plane for the cases where only the 
$CP$ effect (a, d), only the matter effect (b, e), and both the $CP$ 
and the matter effects (c, f) are taken into account. The upper and 
the lower three panels are for L = 250 km and 730 km, respectively.
In (c, f) we also indicate by the long-dashed symbol the line 
$s_{12}^2$ = 0.5 where the matter effect vanishes. 
The remaining parameters are fixed to be the same as in Fig. 4.

\newpage
\vglue -1.3cm
\centerline{(a) $L$ = 250 km}
\centerline{\hskip -1cm
\psfig{file=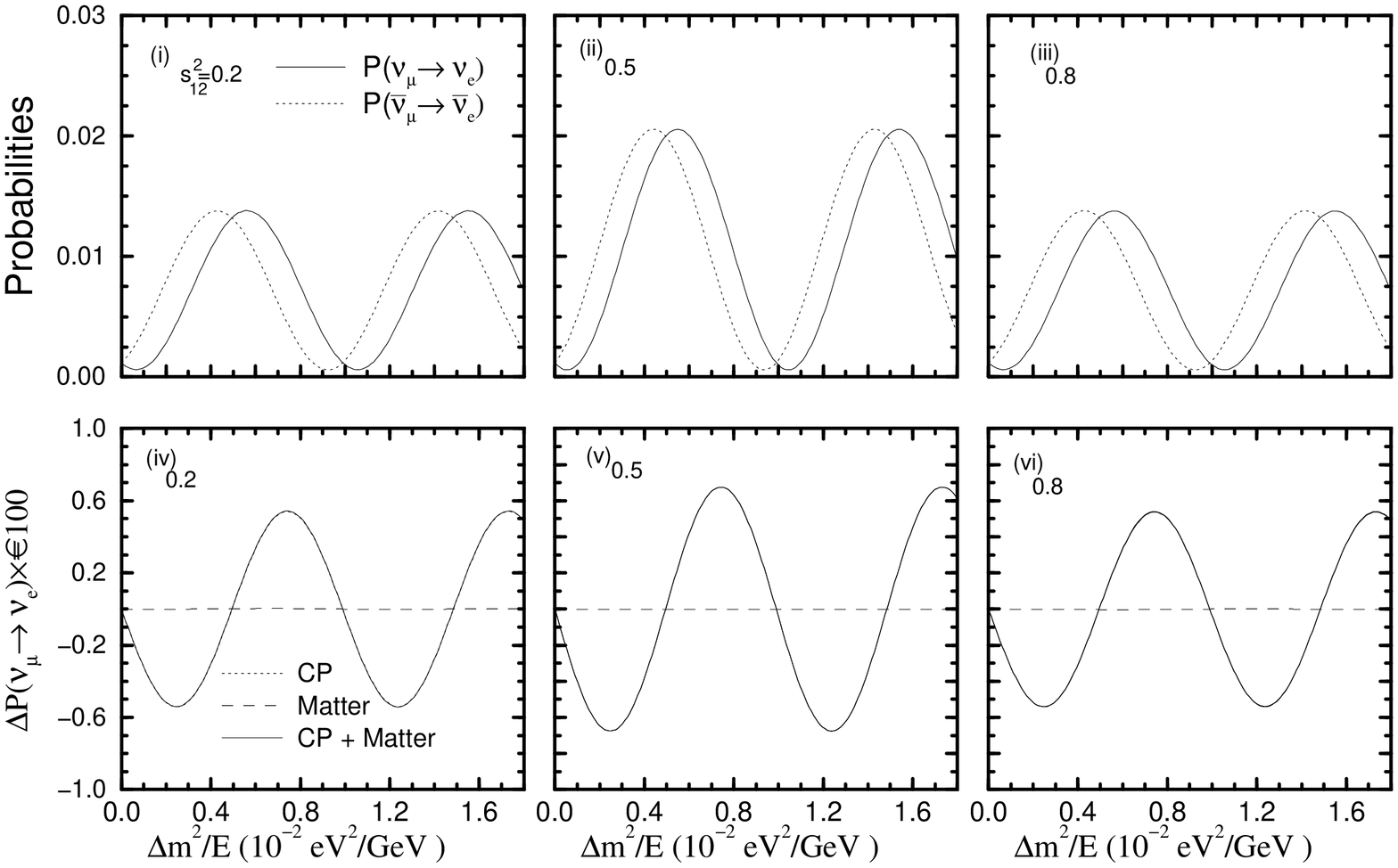,height=11.0cm,width=15.0cm,angle=-90}}
\vglue -3.0cm
\centerline{(b) $L$ = 730 km}
\centerline{\hskip -1cm
\psfig{file=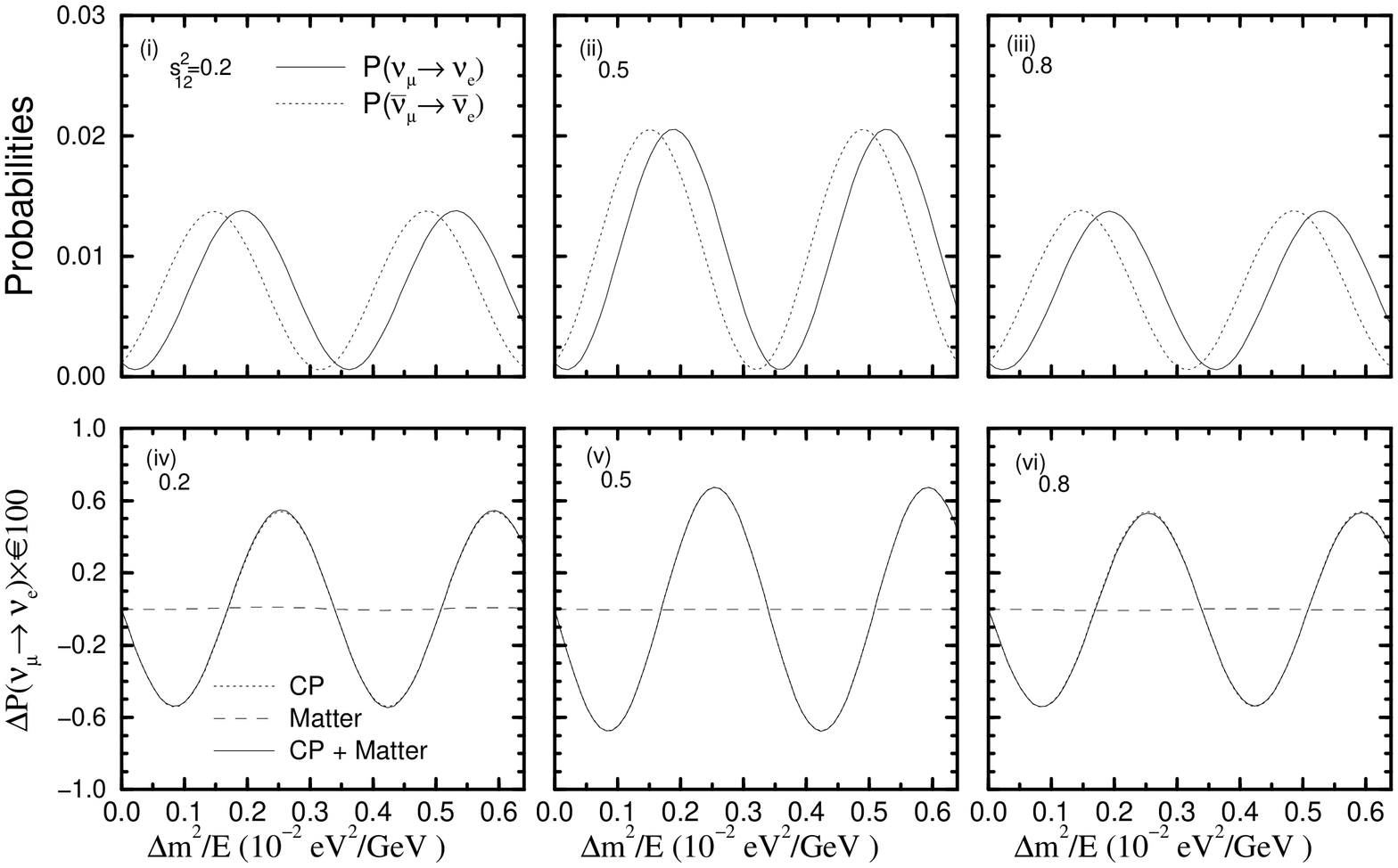,height=11.0cm,width=15.0cm,angle=-90}}
\vglue -2.0cm
\noindent
Fig. 6: Same as in Fig. 4 but for the parameter set
(b) $s_{23}^2 = 2.0\times 10^{-2}$, $s_{13}^2$ = 0.98.

\newpage
\vglue -1.3cm
\centerline{(a) $L$ = 250 km}
\centerline{\hskip -1cm
\psfig{file=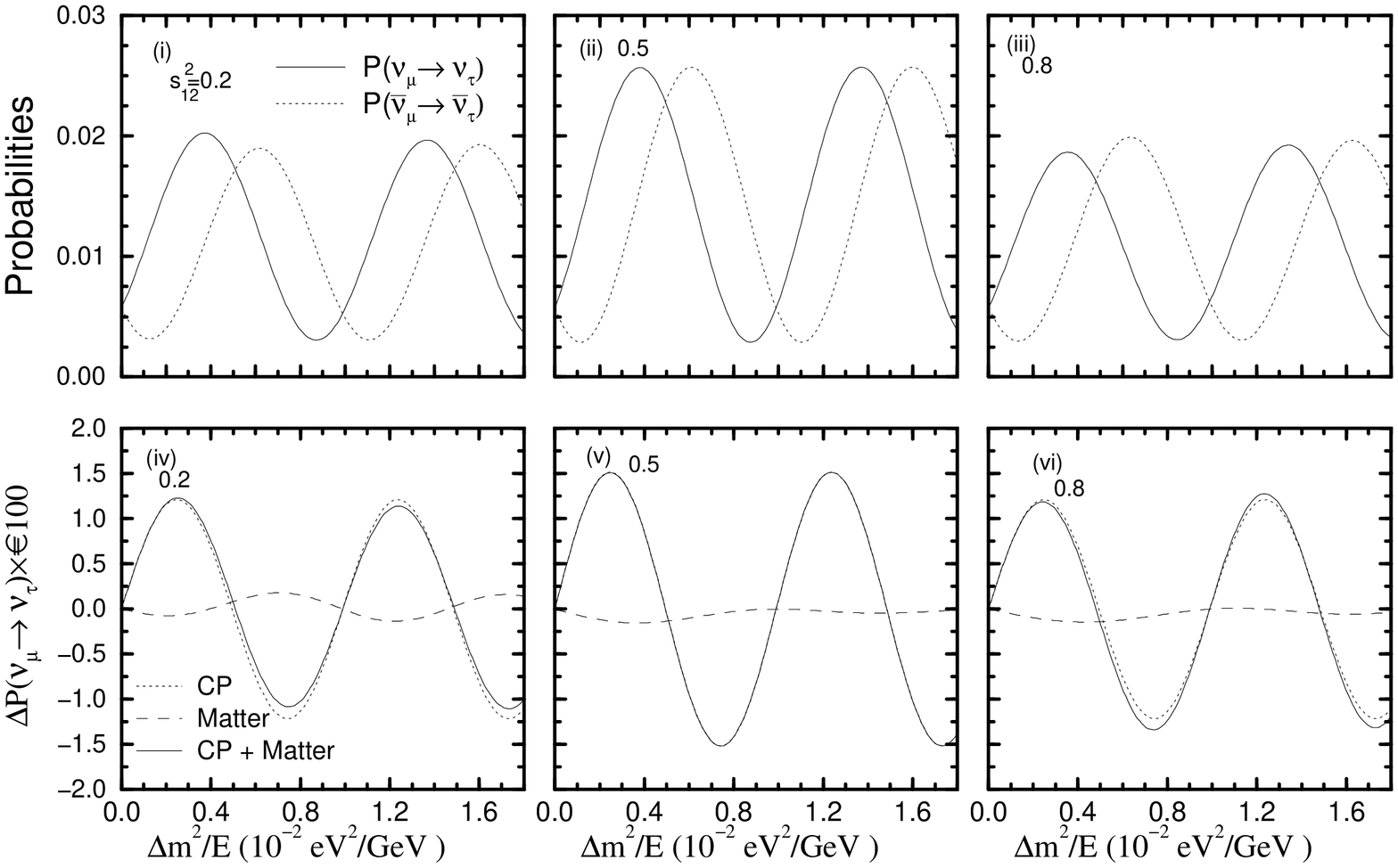,height=11.0cm,width=15.0cm,angle=-90}}
\vglue -3.0cm
\centerline{(b) $L$ = 730 km}
\centerline{\hskip -1cm
\psfig{file=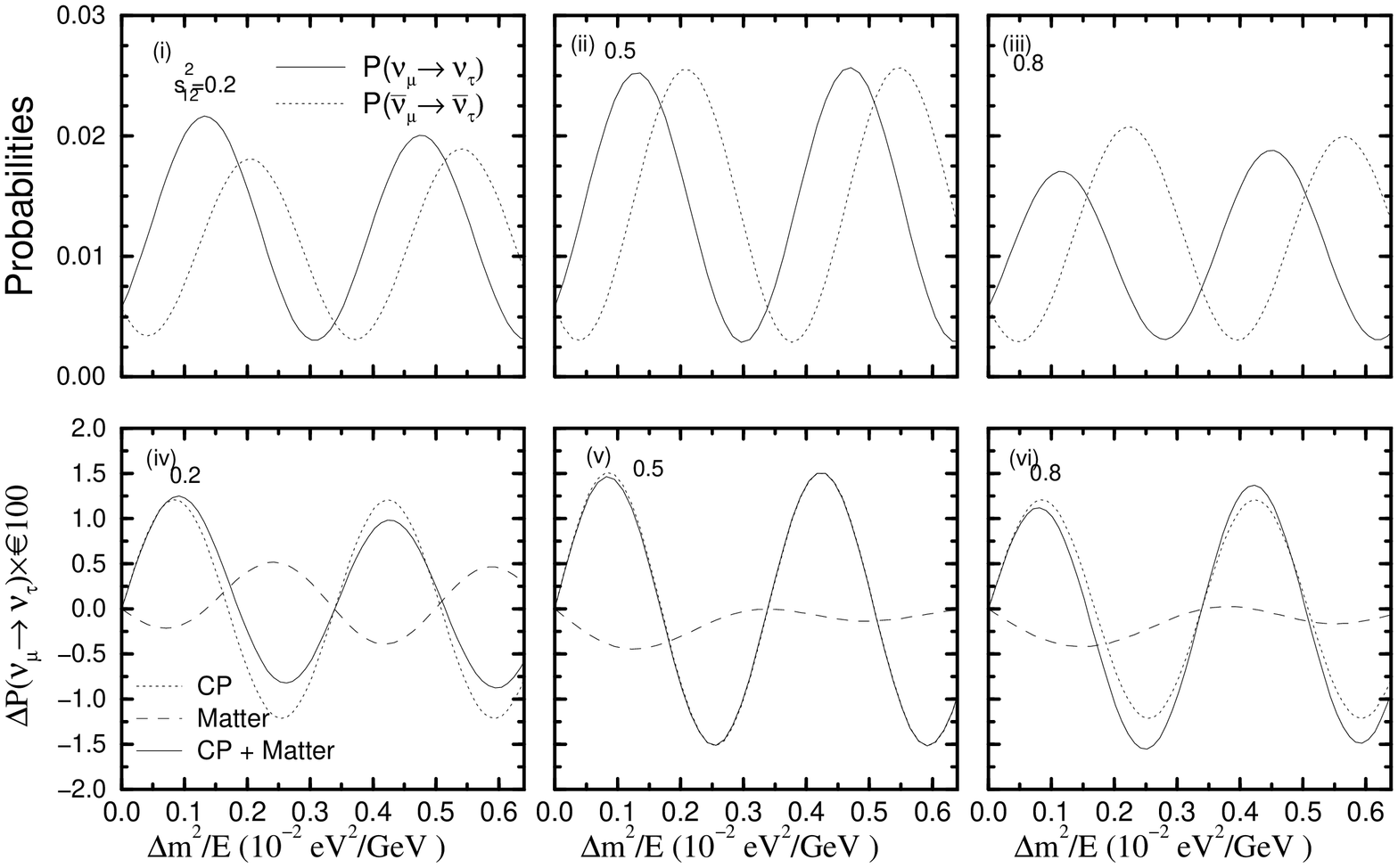,height=11.0cm,width=15.0cm,angle=-90}}
\vglue -2.0cm
\noindent
Fig. 7: Same as in Fig. 4 with the parameter set 
(a) $s_{23}^2 = 3.0\times 10^{-3}$, $s_{13}^2 = 2.0\times 10^{-2}$ 
but for 
$\nu_\mu\to\nu_\tau$ and $\bar{\nu}_\mu\to\bar{\nu}_\tau$ channels.

\newpage
\vglue -1.5cm
\centerline{\hskip -1cm
\psfig{file=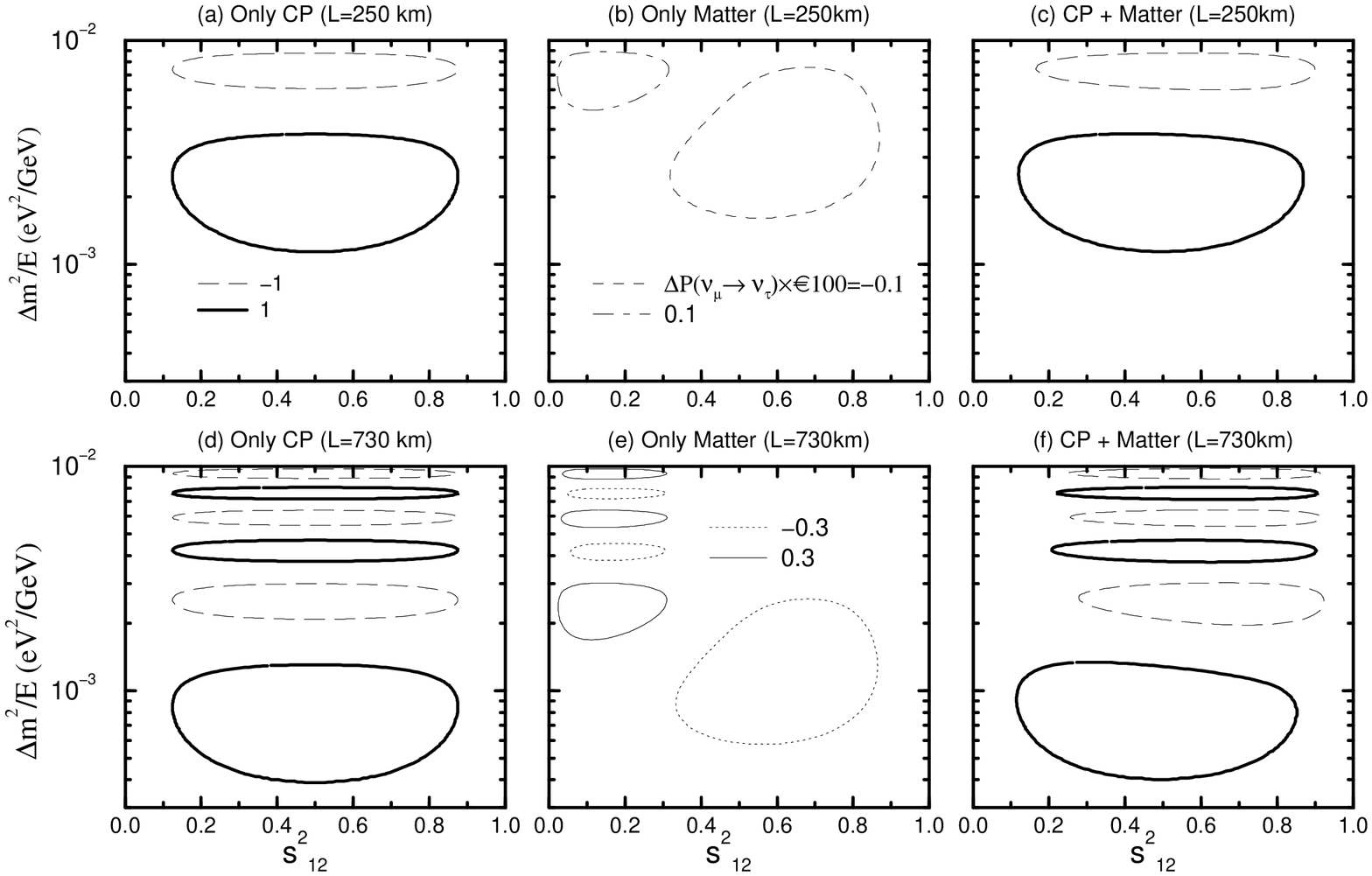,height=18.0cm,width=18.0cm,angle=-90}}
\vglue -1.5cm
\noindent
Fig. 8: Same as Fig. 5 but for $\nu_\mu\to\nu_\tau$ channel with 
the parameter set (a) $s_{23}^2 = 3.0\times 10^{-3}$, 
$s_{13}^2 = 2.0\times 10^{-2}$. 

%
\newpage
\vglue -1.3cm
\centerline{(a) $L$ = 250 km}
\centerline{\hskip -1cm
\psfig{file=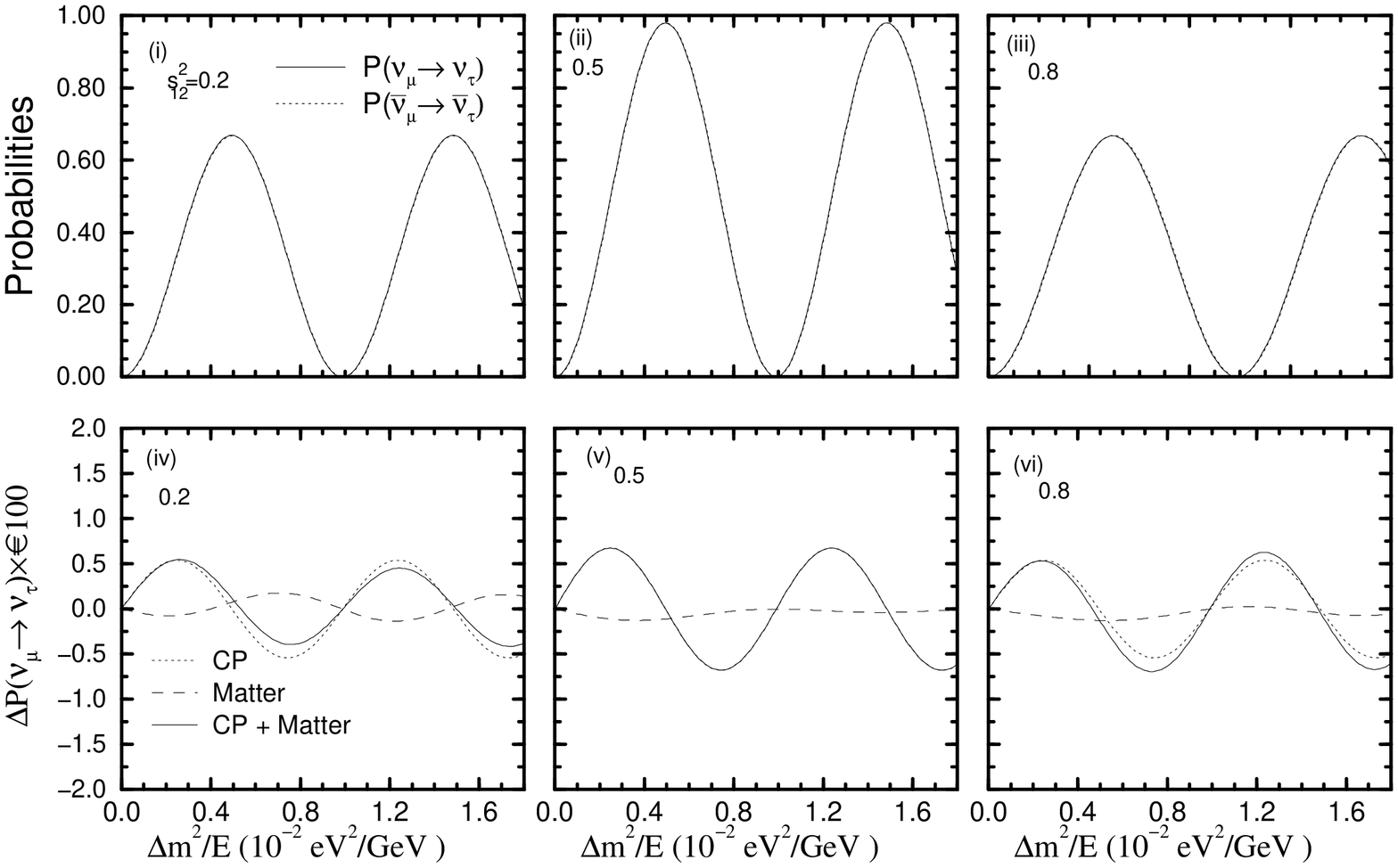,height=11.0cm,width=15.0cm,angle=-90}}
\vglue -3.0cm
\centerline{(b) $L$ = 730 km}
\centerline{\hskip -1cm
\psfig{file=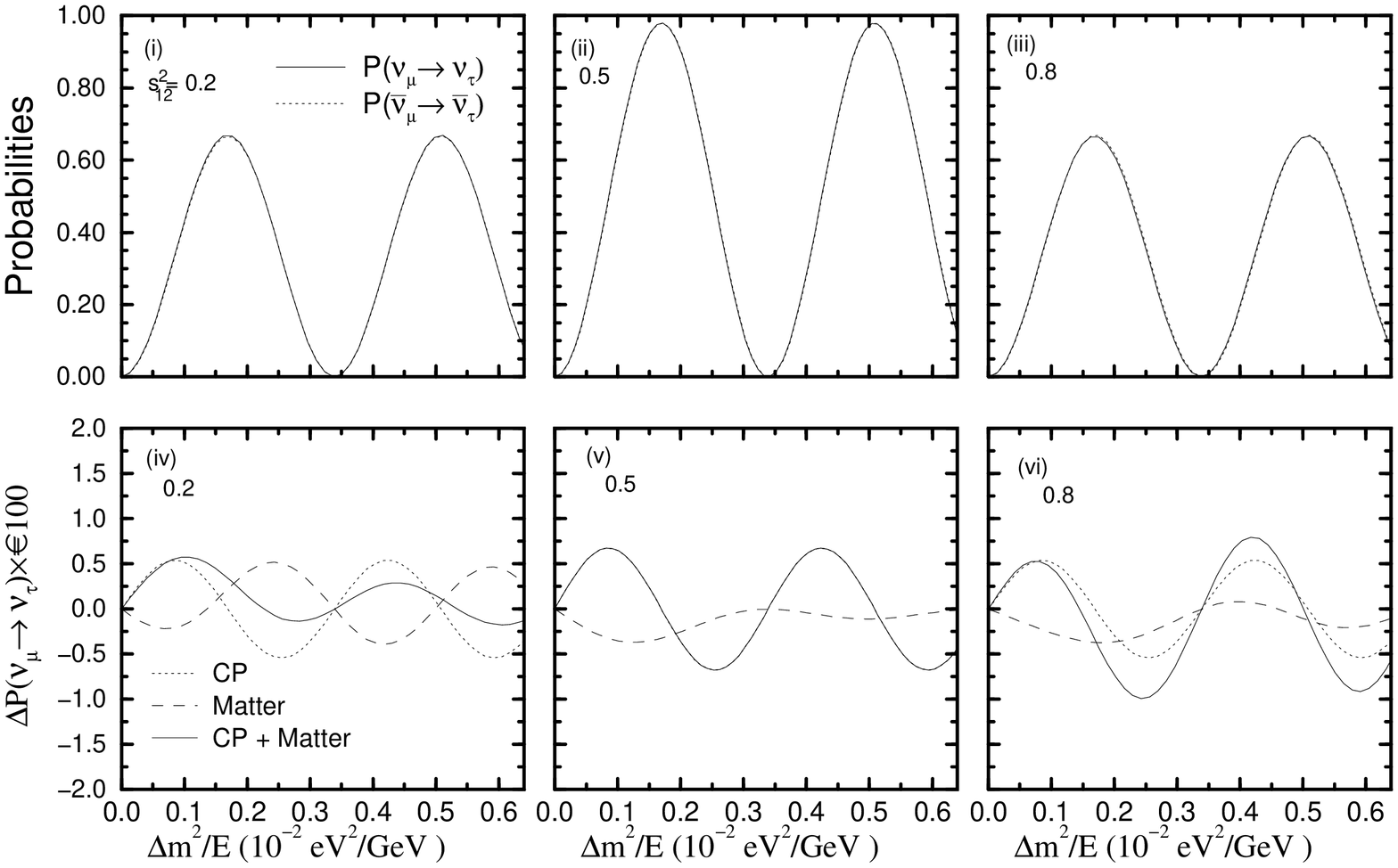,height=11.0cm,width=15.0cm,angle=-90}}
\vglue -2.0cm
\noindent
Fig. 9: Same as in Fig. 7 but for the parameter set
(b) $s_{23}^2 = 2.0\times 10^{-2}$, $s_{13}^2$ = 0.98.

\newpage
\vglue -1.5cm
\centerline{\hskip -1cm
\psfig{file=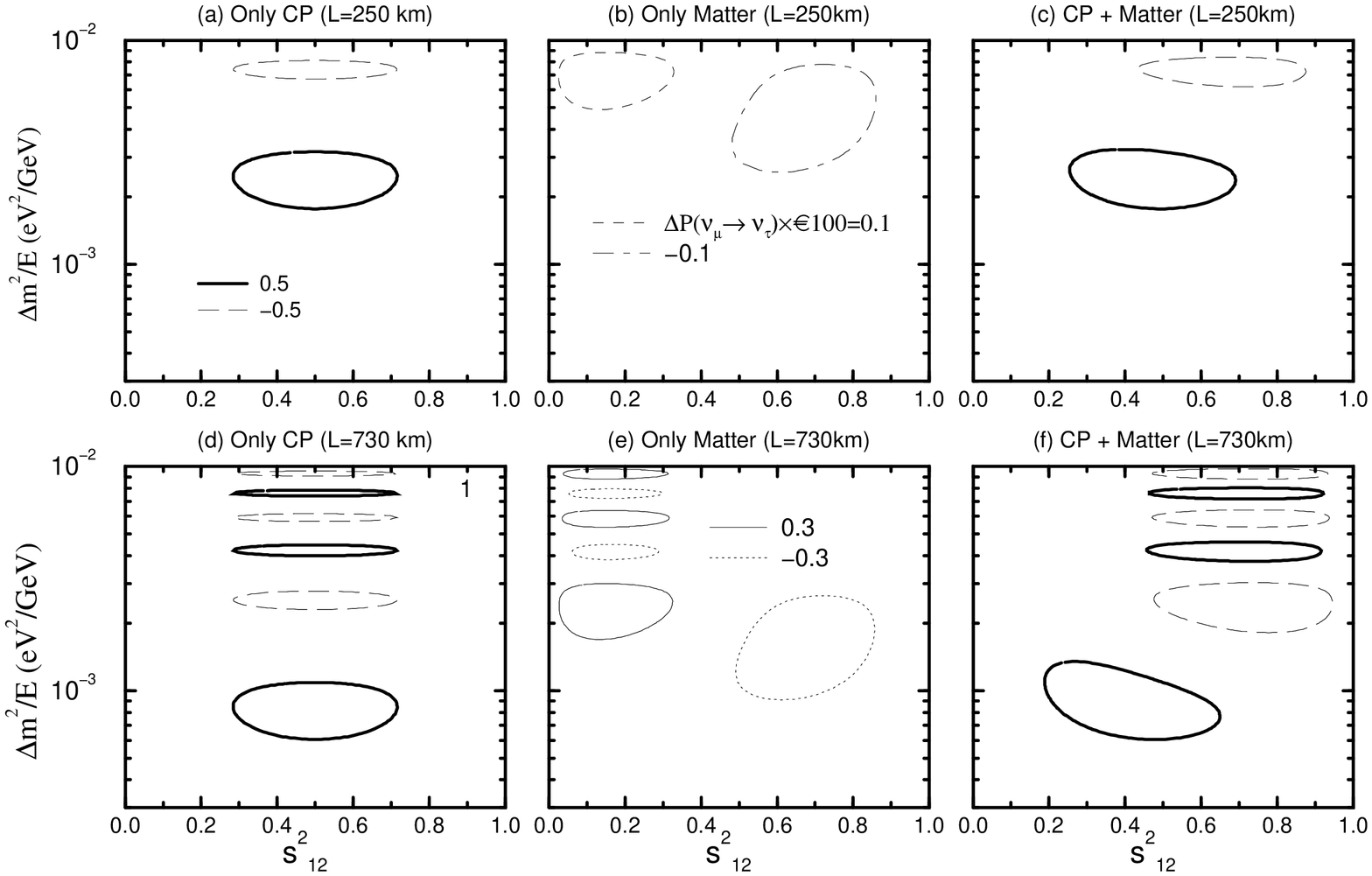,height=18.0cm,width=18.0cm,angle=-90}}
\vglue -1.5cm
\noindent
Fig. 10: Same as Fig. 8 but for the parameter set 
(b) $s_{23}^2 = 2.0\times 10^{-2}$, $s_{13}^2$ = 0.98. 

\end{document}